\documentclass[final,5p,times,twocolumn]{elsarticle}

\usepackage{amsmath,amssymb}
\usepackage{graphicx}
\usepackage{geometry}
\usepackage{array}
\usepackage{multicol}
\usepackage{tabularx}
\usepackage{subcaption}
\usepackage{lipsum}
\usepackage{natbib}
\usepackage{amsmath}
\usepackage{textcomp}     
\usepackage{textgreek}    
\usepackage[table,xcdraw]{xcolor} 

\usepackage[colorlinks=true,
            linkcolor=blue,
            citecolor=blue,
            urlcolor=blue]{hyperref}

\usepackage{footmisc}

\journal{Physics Letters B}
\makeatletter
\def\ps@pprintTitle{%
 \let\@oddhead\@empty
 \let\@evenhead\@empty
 \let\@oddfoot\@empty
 \let\@evenfoot\@empty
}
\makeatother

\begin{document}

\begin{frontmatter}

\author[1]{Abhishek\corref{cor1}}
\ead{abphy96@gmail.com}

\cortext[cor1]{Corresponding author}

\title{Low scale leptogenesis and TM\textsubscript{1} mixing in neutrinophilic two Higgs doublet model (\textnu2HDM) with S\textsubscript{4} flavor symmetry}

\author{V. Suryanarayana Mummidi\textsuperscript{1,\textdagger}}
\address{\textsuperscript{1}Department of Physics, National Institute of Technology, Tiruchirappalli 620015, Tamil Nadu, India \vspace{-0.45 in}}

\begin{abstract}
We study a modified version of the Standard Model that includes a scalar doublet and two right-handed neutrinos, forming the neutrinophillic two higgs doublet ($\nu\text{2HDM}$) framework. For the two Higgs-doublet vacuum expectation values satisfying $\text{v}_2 << \text{v}_1$, this model operates at a TeV scale, bringing the RHNs within experimental reach. To further enhance its predictive power, we introduce an $S_4 \times Z_4$ flavor symmetry with five flavons, resulting in mass matrices that realize the so-called Trimaximal $\text{TM}_1$ mixing scheme. The model effectively explains lepton masses and flavor mixing under the normal ordering of neutrino masses, predicting that the effective neutrino mass in 0$\nu\beta\beta$ decay lies between [4 - 5] meV, significantly lower than the sensitivity limits of current experiments. We also investigate how this framework could support low-scale leptogenesis as a natural way to explain the observed imbalance between matter and antimatter in the Universe. This work explores how neutrino physics, flavor symmetry and baryon asymmetry are connected, providing a clear framework that links theoretical predictions with experimental possibilities.
\end{abstract}

\end{frontmatter}

\section{Introduction}
\label{introduction}

The Standard Model (SM) has significantly contributed to describing fundamental particle interactions, yet it does not fully explain several profound questions in physics. These include the unexplained small masses of neutrinos \cite{deSalas:2017kay, PDG2020}, the puzzling imbalance between matter and antimatter in the cosmos—known as baryon asymmetry (BAU) ~\cite{buchmuller2005}—and the elusive nature of dark matter (DM) ~\cite{ma2006}. These gaps underscore the need to investigate theories that go beyond the Standard Model (BSM).

One of the significant advancements in particle physics was the discovery of neutrino oscillations~\cite{fukuda1998,ahmad2001}, which provided strong evidence that neutrinos possess a small yet nonzero mass. This finding necessitates a robust theoretical framework to elucidate the origin of neutrino masses. A well-established approach is the Type-I seesaw mechanism, which introduces right-handed neutrinos (RHNs) as singlets under the SM gauge symmetry. In this framework, neutrino masses emerge via the seesaw relation $m_{\nu} \sim \frac{\text{v}^2 y^2}{m_N},$ where \( y \) represents the Yukawa coupling, \( \text{v} \) denotes the Higgs field’s vacuum expectation value (vev) and \( m_N \) corresponds to the mass scale of RHNs~\cite{minkowski1977, Mohapatra1980}. The suppression factor \( m_N \) naturally accounts for the smallness of neutrino masses, even when the Yukawa coupling is of order \( \mathcal{O}(1) \), provided that \( m_N \) is sufficiently large.

Another positive outcome of the seesaw mechanism, is that it offers a compelling explanation for the universe’s matter-antimatter asymmetry through leptogenesis. However, a key challenge arises when considering the scale of \( m_N \) required for successful leptogenesis. In traditional high-scale seesaw models, RHNs typically acquire masses in the range of $10^9$ to $10^{15}$ GeV, making it challenging to probe such models in near-future experiments. However, the large separation between this high mass scale and the electroweak scale (\({\sim}10^2\) GeV) introduces a hierarchy problem, raising questions about the naturalness of such models. Such a disparity raises concerns about fine-tuning and suggests that additional mechanisms, such as radiative corrections or symmetry-based protections, may be necessary to stabilize the neutrino mass parameters~\cite{Vissani1998}.  

To overcome these challenges, alternative pathways for successful low-scale leptogenesis have been explored. Among these, resonant leptogenesis~\cite{Pilaftsis2004} exploits a nearly degenerate spectrum of RHNs, enhancing CP violation through self-energy corrections. Akmedov-Rubakov-Smirnov (ARS) leptogenesis~\cite{akhmedov1998,Asaka2005} operates at the GeV scale, where RHN flavor oscillations generate the required lepton asymmetry. Another viable approach involves Higgs decays~\cite{Hambye2016,Hambye2017}, where CP-violating interactions in scalar decays contribute to the production of the observed BAU. However, many of these mechanisms rely on a highly fine-tuned RHN mass spectrum to maintain the required CP asymmetry while avoiding excessive washout effects~\cite{barry2010}. 

To reconcile the seesaw framework with low-scale phenomena while avoiding fine-tuning, several BSM extensions have been proposed. The \(\nu2\)HDM~\cite{Ma2001,Sarma:2022qka} builds upon the SM by incorporating an additional Higgs doublet. This extension provides a natural mechanism for generating neutrino masses while also influencing the dynamics of leptogenesis. Another compelling alternative is the scotogenic model~\cite{ma2006, Sarma2020, Borah2019}, in which neutrino masses arise radiatively through loop corrections, with additional dark-sector particles playing a crucial role. These low-scale extensions provide experimentally testable predictions, making them attractive candidates for future collider searches, neutrino experiments, and cosmological observations. 

The exploration of neutrino flavor mixing has been a central focus in particle physics, with various theoretical frameworks proposed to explain the observed mixing patterns. Among these, the tri-bimaximal (TBM)\cite{Harrison:2002er} mixing scheme was initially considered a strong candidate. However, experimental results from Daya Bay\cite{An:2012eh}, RENO\cite{Ahn:2012nd}, and Double Chooz\cite{Abe:2012tg} confirmed that TBM requires modifications to accommodate a nonzero reactor mixing angle ({$\theta_{13} \neq 0$}). One well-justified approach to explaining these deviations is the Trimaximal $\text{TM}_1$ mixing scheme~\cite{Luhn:2013fh, Grimus:2013pg, Rodejohann:2012kx,Ganguly:2023jml}, which preserves the leading column of the TBM mixing matrix while introducing essential modifications via a 2-3 rotation.

Among discrete symmetries, $S_4$ has gained prominence due to its ability to naturally produce trimaximal $\text{TM}_1$ mixing, a flavor structure consistent with global neutrino oscillation data, including nonzero $\theta_{13}$~\cite{lam2008,altarelli2010}. The mathematical elegance of $S_4$ symmetry allows for hierarchical RHN masses while simultaneously enabling leptogenesis at the TeV scale~\cite{antusch2004}. This unification of neutrino mass generation and BAU makes $S_4$-based $\nu2$HDM frameworks particularly compelling.

Within this framework, we have explicitly determined the Yukawa couplings, showing their consistency with experimental data on neutrino oscillations and the constraints imposed by lepton flavor violation (LFV) processes. This study extends the $\nu2$HDM model by incorporating an $S_4 \times Z_4$ symmetry. Furthermore, we have analyzed the rate of neutrinoless double-beta decay ($0\nu\beta\beta$), which serves as a crucial probe of the Majorana properties of neutrinos and helps impose limitations on the model parameters. Additionally, we have conducted a thorough investigation of leptogenesis, in which the rotational angle, Yukawa interactions, and the RHN mass spectrum all have a sensitive effect on the generated lepton asymmetry. By exploring different choices of this angle, we have analyzed its impact on neutrino phenomenology, cosmological studies and LFV observables, highlighting how specific model parameters shape testable predictions for future experimental verification. 

We have divided the paper into the following seven sections. Sec.~\ref{Sec:model} presents the construction of the model incorporating flavor discrete symmetry and discusses its key characteristics. Sec.~\ref{sec:MA} defines the model parameters and predicts the neutrino observables, ensuring consistency with neutrino oscillation data within the \(3\sigma\) range. Sec.~\ref{Sec:lFV} is dedicated to the calculation of LFV constraints, while Sec.~\ref{Sec:lept} focuses on the analysis of leptogenesis. In Sec.~\ref{Sec:NA}, we present the results of our leptogenesis study, followed by Sec.~\ref{Sec:conclusion}, which provides the conclusions of our work.

\section{Model}
\label{Sec:model}

\begin{table*}[t] 
\centering
\renewcommand{\arraystretch}{1.2} 
\setlength{\tabcolsep}{10pt} 
\caption{Field assignments under \( S_4 \otimes Z_4 \).}
\begin{tabular}{c|c c c c c c c c c c c c c c}
\hline
\(Field\) & \(\ell_L\) & \(e_R\) & \((\mu_R, \tau_R)\) & \(\phi_1\) & \(\phi_2\)  & \(N_1\)& \(N_2\) & \(\chi\) & \(\chi'\) & \(\psi\) & \(\rho\) & \(\rho'\) & \(\phi_\ell\) & \(\Phi_\ell\) \\ \hline
\(SU(2)\)   & \(2\)& \(1\) & \(1\)  & \(2\) & \(2\) & \(1\) & \(1\) & \(1\) & \(1\) & \(1\) & \(1\) & \(1\) & \(1\) & \(1\) \\ \hline
\(S_4\)     & \({3_1}\) & \({1_1}\) & \({2}\) & \({1_1}\) & \({1_2}\)  & \({1_2}\) & \({1_1}\) & \({3_1}\) & \({3_2}\) & \({3_2}\)& \({1_1}\)& \({1_1}\) & \({3_1}\) & \({3_2}\) \\ \hline
\(Z_4\)   & \(i\)& \(1\) & \(1\)  & \(1\) & \(1\) & \(-1\) & \(-i\) & \(i\) & \(1\) & \(1\) & \(1\) & \(-1\) & \(-i\) & \(-i\) \\ \hline
\end{tabular}
\label{tab:discrete_symmetry}
\end{table*}

The $\nu$2HDM~\cite{ma2006} extends the SM by introducing two massive RHNs along with an additional scalar doublet ($\phi_2$), vev obtained in this case is significantly suppressed. While the conventional version of this model generally incorporates three RHNs, our study explores a minimal extension featuring only two RHNs. A fundamental feature of the $\nu$2HDM is the incorporation of a global $U(1)_L$ symmetry, which enforces specific charge assignments: $L_{{\phi}_1} = 0$, $L_{{\phi}_2} = -1$, and $L_N = 0$. Because of this symmetry, the extra scalar doublet $\phi_2$ links only to heavy neutrinos. However, the SM Higgs doublet (${\phi}_1$) continues to interact with quarks and charged leptons in a way consistent with the SM. In this scenario the neutrino mass  can be understood using the dimension-five Weinberg operator \cite{Weinberg1979,Ma1998}:
\begin{equation}
\frac{(\nu_i \Phi^0 - l_i \Phi^+)(\nu_j \Phi^0 - l_j \Phi^+)}{\Lambda},
\end{equation}

For an effective large mass scale, $\Lambda$ is used. This operator is essential to the production of light neutrino masses in the seesaw process.  Consequently, the light neutrino's mass is determined by \cite{Mohapatra1980}:

\begin{equation}
m_{\nu} \approx \frac{m_D^2}{m_N},
\end{equation}

In this case, $m_N = \Lambda f$ and $m_D = f \text{v}$ represent the mass of RHN and Dirac mass of the neutrino respectively, here $\text{v}$ corresponds to the Higgs field’s vev. In traditional seesaw framework, $m_N$ is extremely large, making direct experimental tests challenging. However, in $\nu$2HDM, the mass scale of $m_N$ can be around 1 TeV, allowing experimental verification.

The $\nu$2HDM offers an elegant and experimentally testable framework for neutrino mass generation. Tiny mass of neutrino's arises naturally out of the suppressed vev of $\phi_2$, avoiding the need for too much fine-tuning. Based on the charge assignments under the $SU(3)_C \otimes SU(2)_L \otimes U(1)_Y$ gauge group, the fundamental particle spectrum of the $\nu$2HDM model consists of:

\begin{align}
    \begin{bmatrix}
\nu_i  \\
\ell_i
\end{bmatrix}_L & = (1, 2, -1/2), & l_{iR} & = (1, 1, -1), & N_i & = (1, 1, 0), \\
    \phi_1 & = (1, 2, 1/2), & \phi_2 & = (1, 2, 1/2).
\end{align}

The following is an expression for the model's scalar doublets, $\phi_1$ and $\phi_2$:
\begin{align}
    \phi_1 = \begin{pmatrix}
        \phi_1^+ \\
        \frac{\text{v}_1 + \phi_1^{0,r} + i \phi_1^{0,i}}{\sqrt{2}}
    \end{pmatrix}, \quad
    \phi_2 = \begin{pmatrix}
        \phi_2^+ \\
        \frac{\text{v}_2 + \phi_2^{0,r} + i \phi_2^{0,i}}{\sqrt{2}}
    \end{pmatrix}.
\end{align}
Here, $\text{v}_1$ and $\text{v}_2$ denote the vev's associated with the two scalar doublets. The corresponding Higgs potential is given by:
\begin{equation}
\begin{aligned}
    V = & \ m_{\phi_1}^2 \phi_1^\dagger \phi_1 + m_{\phi_2}^2 \phi_2^\dagger \phi_2  
    + \frac{\lambda_1}{2} (\phi_1^\dagger \phi_1)^2 + \frac{\lambda_2}{2} (\phi_2^\dagger \phi_2)^2 \\
    & + \lambda_3 (\phi_1^\dagger \phi_1)(\phi_2^\dagger \phi_2) 
    + \lambda_4 (\phi_1^\dagger \phi_2)(\phi_2^\dagger \phi_1) 
    \\
    & - \mu^2 \zeta (\phi_1^\dagger \phi_2 + \text{h.c.})  + \left(\frac{\lambda_5}{2} (\phi_1^\dagger \phi_2)^2 + \text{h.c.}\right).
\end{aligned}
\label{Eq:4}
\end{equation}
The Higgs potential given in Eq.~\eqref{Eq:4} explains how the scalars, $\phi_1$ and $\phi_2$, interact. $m_{\phi_1}$ and $m_{\phi_2}$ represent the mass terms for the scalar doublets, while the $\lambda_i$ describe their self-interactions and interactions between the two doublets. The $\mu_{12}$ term explicitly and softly breaks the lepton symmetry. $\mu_{12}$ is defined as $\mu_{12} = \mu \text{v}_\zeta$, where $\text{v}_\zeta$ stands for $\zeta$'s vev. The following criteria ensure that the above potential has a stable vacuum and is bounded from below~\cite{Gunion:2002zf}: 
\begin{align}
\lambda_1 > 0, \,   
\lambda_2 > 0, \,
\lambda_3 > -\sqrt{\lambda_1 \lambda_2}, \,
\lambda_3 + \lambda_4 - |\lambda_5| > -\sqrt{\lambda_1 \lambda_2}
\end{align}

The minimization condition is given by the following equations:
\begin{align}
    m_{\phi_1}^2 & = \mu_{12}^2 \frac{\text{v}_2}{\text{v}_1} - \frac{\lambda_1}{2} \text{v}_1^2 - \frac{\lambda_3 + \lambda_4 + \lambda_5}{2} \text{v}_2^2 \\
    m_{\phi_2}^2 & = \mu_{12}^2 \frac{\text{v}_1}{\text{v}_2} - \frac{\lambda_2}{2} \text{v}_2^2 - \frac{\lambda_3 + \lambda_4 + \lambda_5}{2} \text{v}_1^2
\end{align}

This helps to express the vev of the scalar doublets in terms of the parameters of the Higgs potential. As the only source of lepton number violation, radiative corrections to the term $\mu_{12}^2$ are proportional to $\mu_{12}^2$ itself and logarithmically sensitive to the cutoff scale~\cite{DavidsonLogan2009}. Stabilizing the vev hierarchy $\text{v}_2 \ll \text{v}_1$ against radiative corrections is achieved in this way.  The physical Higgs bosons following spontaneous symmetry breaking (SSB) are determined by:
\begin{align}
    H^+ = \phi_2^+ \cos\beta - \phi_1^+ \sin\beta, \quad A  = \phi_2^{0,i} \cos\beta - \phi_1^{0,i} \sin\beta \\
    H^0  = \phi_2^{0,r} \cos\alpha - \phi_1^{0,r} \sin\alpha, \quad h  = \phi_2^{0,r} \cos\alpha + \phi_1^{0,r} \sin\alpha
\end{align}

 The Higgs spectrum consists of a positively charged Higgs boson (\( H^+ \)), a CP-odd scalar (\( A \)), a heavier CP-even Higgs boson (\( H^0 \)) and a lighter CP-even Higgs boson (\( h \)). Their composition is determined by the mixing angles \( \beta \) and \( \alpha \), which are given by:
\begin{align}
   \tan\beta = \frac{\text{v}_1}{\text{v}_2},\quad \tan2\alpha \simeq 2\frac{\text{v}_2}{\text{v}_1} \frac{-\mu_{12}^2+ (\lambda_3 + \lambda_4 + \lambda_5)\text{v}_1\text{v}_2}{-\mu_{12}^2 + \lambda_1 \text{v}_1\text{v}_2}.
\end{align}

after the omission of $\mathcal{O}(\text{v}^2)$ and $\mathcal{O}(\mu_{12}^2)$, the approximate mass expressions for the physical Higgs bosons are given by:
\begin{equation}
m_{H^+}^2 \simeq m_{\phi_2}^2 + \frac{\lambda_3}{2} \text{v}_1^2,\quad
m_A^2 \simeq m_{H^+}^2 + \frac{\lambda_4 - \lambda_5}{2} \text{v}_1^2
\end{equation}
\begin{equation}
 m_{H^0}^2 \simeq m_{H^+}^2 + \frac{\lambda_4 + \lambda_5}{2} \text{v}_1^2,\quad
m_h^2 \simeq \lambda_1 \text{v}_1^2.
\end{equation}

The model's Higgs sector's hierarchical structure is emphasized by these approximations of mass relations. The lighter CP-even Higgs boson (\( h \)) is primarily governed by the quartic coupling \( \lambda_1 \), whereas the heavier Higgs states (\( H^+ \), \( A \), and \( H \)) acquire their masses through contributions from the couplings \( \lambda_3 \), \( \lambda_4 \) and \( \lambda_5 \).

In the \( \nu2 \)HDM, the dynamics of neutrino mass generation can be described by the following effective Lagrangian: 

\begin{equation}
    \mathcal{L} \subset Y \bar{l}_L \tilde{\phi}N + \frac{1}{2} \bar{N}^c m_N N + \text{h.c.}
    \label{Eq:13}
\end{equation}

The left-handed lepton doublet is represented by \( l_L \), while \( N \) denotes the RHNs. The conjugate Higgs doublet is defined as \( \tilde{\phi} = i\sigma_2 \phi^* \). To account for the observed patterns in the lepton sector, we introduce an \( S_4 \otimes Z_4 \) flavor symmetry within the \( \nu2 \)HDM framework.
 The mass matrices structure is constrained by this symmetry, which also yields certain predictions for the mass hierarchy and neutrino mixing angles.  Table~\ref{tab:discrete_symmetry} provides the model's particle composition and charge assignments under $S_4 \otimes Z_4$. In addition to the SM fields, we introduce flavon fields $\chi$, $\chi'$, $\psi$, $\rho$ and $\rho'$, which are responsible for breaking the $S_4 \otimes Z_4$ symmetry and generating the desired mass structures.

Yukawa Lagrangian for lepton sector is given by:
\begin{equation}
   \mathcal{L}_{\ell} = \frac{Y_{l_1}}{\Lambda} \bar{\ell_L}\phi_1 \phi_\ell e_R + \frac{Y_{l_2}}{\Lambda} \bar{\ell_L}\phi_1 \phi_\ell (\mu_R, \tau_R) +\frac{Y_{l_3}}{\Lambda} \bar{\ell_L}\phi_1 \Phi_\ell (\mu_R, \tau_R) + h.c
\end{equation}
Here, \( Y_{l_1} \), \( Y_{l_2} \) and \( Y_{l_3} \) denote the Yukawa coupling constants for the electron, muon, and tau sectors, respectively, while \( e_R \), \( \mu_R \) and \( \tau_R \) are the corresponding right-handed charged leptons.
 
We introduce flavon fields \( \phi_\ell \) and \( \Phi_\ell \), which acquire vev's given by \( \langle \phi_\ell \rangle = (\text{v}_{\phi_\ell}, 0, 0) \) and \( \langle \Phi_\ell \rangle = (\text{v}_{\Phi_\ell}, 0, 0) \), respectively. The flavons \( \phi_\ell \) and \( \Phi_\ell \) are assigned under \( S_4 \) to yield a diagonal charged lepton mass matrix after spontaneous symmetry breaking~\cite{Zhao:2011}.

\begin{equation}
    m_\ell=\frac{\langle\phi_1\rangle}{\Lambda}
    \begin{bmatrix}
         Y_{l_1} \text{v}_{\phi_\ell}&0 &0\\
         0& Y_{l_2} \text{v}_{\phi_\ell} + Y_{l_3} \text{v}_{\Phi_\ell} & 0\\
         0 &0& Y_{l_2} \text{v}_{\phi_\ell} - Y_{l_3} \text{v}_{\Phi_\ell}
    \end{bmatrix}
\end{equation}

In the neutrino sector, the Dirac mass matrix arises from the interactions between the Higgs doublet \( \phi_2 \), left-handed neutrinos \( \nu_L \) and RHNs \( N \) via Yukawa couplings. The corresponding Lagrangian for Dirac mass terms can be written as:

\begin{equation}
 \mathcal{L}_D =  \frac{Y_1}{\Lambda}(\ell_L \tilde{\phi}_2 \chi)N_1 +  \frac{Y_2}{\Lambda}(\ell_L \tilde{\phi}_2 \chi')N_2 +  \frac{Y_3}{\Lambda}(\ell_L \tilde{\phi}_2\psi)N_2 + h.c
\end{equation}

The Majorana mass terms associated with the RHNs are given by:

\begin{equation}
 \mathcal{L}_R = Y_{N_1} \bar{N_1}^c N_1 \rho + Y_{N_2} \bar{N_2}^c N_2 \rho' + h.c
\end{equation}

The observed hierarchy in the charged-lepton mass spectrum is accounted for through the Froggatt–Nielsen (FN) mechanism, as implemented following the framework detailed in~\cite{Zhao:2011}. The flavon fields and their alignments are chosen as follows:
\begin{align}
    \langle\chi\rangle = (w,w,w),\quad \langle\chi'\rangle =
    (0,w,-w), \\
    \langle\psi\rangle = (w,w,w),\quad \langle\rho\rangle= \langle\rho'\rangle= w.
\end{align}

Here, $Y_e$, $Y_\mu$ and $Y_\tau$ denote the Yukawa couplings corresponding to the electron, muon and tau, respectively. The vev of \( \psi \) satisfies orthogonality conditions: \( \langle \psi \rangle \cdot \langle \phi_\ell \rangle = 0 \), \( \langle \psi \rangle \cdot \langle \Phi_\ell \rangle = 0 \) and \( \langle \psi \rangle \cdot \langle \chi' \rangle = 0 \).

The Dirac and Majorana mass matrices after symmetry breaking takes the form:
\begin{align}
    m_D= \frac{\langle\phi_2\rangle w}{\Lambda}
    \begin{bmatrix}
         Y_1 & Y_3\\
          Y_1 & -Y_2+Y_3\\
         Y_1 & Y_2+Y_3
    \end{bmatrix},\quad
    m_R=
    \begin{bmatrix}
        M_1  & 0\\
        0    & M_2
    \end{bmatrix}
\end{align}
The Majorana masses for $N_1$ and $N_2$ are denoted by $M_1$ and $M_2$, respectively.
Eq.~\eqref{Eq:13} provides the mass matrix for light neutrinos, which aligns with the type-I seesaw mechanism~\cite{minkowski1977}:  

\begin{equation}
    m_\nu = -\frac{\text{v}_2^2}{2} Y M_N^{-1} Y^T = U_{\text{PMNS}} m_{\nu}^{\text{diag}} U_{\text{PMNS}}^T,
    \label{eq:20}
\end{equation}

where the diagonal neutrino mass matrix is expressed as $\text{diag}(m_1, m_2, m_3) = m_{\nu}^{\text{diag}}$, and $U_{\text{PMNS}}$ represents the Pontecorvo Maki Nakagawa Sakata (PMNS) mixing matrix. The resulting neutrino mass matrix $m_{\nu}$ is structured as follows:

\begin{align}
    m_\nu = \frac{-\text{v}_2^2w^2}{2\Lambda^2} \resizebox{0.38\textwidth}{!}{$\begin{bmatrix}
        \frac{{Y_1}^2}{M_1} +  \frac{{Y_3}^2}{M_2}& \frac{{Y_1}^2}{M_1} +  \frac{Y_3(-Y_2+Y_3)}{M_2}&\frac{{Y_1}^2}{M_1} +  \frac{Y_3(Y_2+Y_3)}{M_2}\\
         \frac{{Y_1}^2}{M_1}+\frac{Y_3(-Y_2+Y_3)}{M_2}& \frac{{Y_1}^2}{M_1}+\frac{(-Y_2+Y_3)^2}{M_2}& \frac{{Y_1}^2}{M_1}+\frac{(Y_2+Y_3)(-Y_2+Y_3)}{M_2}\\
         \frac{{Y_1}^2}{M_1} +  \frac{Y_3(Y_2+Y_3)}{M_2}&\frac{{Y_1}^2}{M_1}+\frac{(Y_2+Y_3)(-Y_2+Y_3)}{M_2}&\frac{{Y_1}^2}{M_1}+\frac{(Y_2+Y_3)^2}{M_2}
    \end{bmatrix}$}
\end{align}
This matrix can be expressed in a more compact form as:
\begin{equation}
    m_\nu =  \begin{bmatrix}
        a+c&a-d+c&a+d+c\\
         a-d+c&a+b+c-2d& a-b+c\\
        a+d+c&a+c-b&a+b+c+2d
    \end{bmatrix}
    \label{Eq:22}
\end{equation}
where the definitions of the parameters $a$, $b$, $c$ and $d$ are as follows:
\begin{align}
a = \frac{-\text{v}_2^2w^2}{2\Lambda^2}\frac{{Y_1}^2}{M_1},\quad b = \frac{-\text{v}_2^2w^2}{2\Lambda^2}\frac{{Y_2}^2}{M_2},
\\
 c = \frac{-\text{v}_2^2w^2}{2\Lambda^2}\frac{{Y_3}^2}{M_2},\quad d = \frac{-\text{v}_2^2w^2}{2\Lambda^2}\frac{{Y_2}{Y_3}}{M_2}.  
\end{align}
The neutrino mass matrix $m_\nu$, is diagonalized using the neutrino mixing matrix $U_{\nu}$, which is formulated as:  
\begin{equation}
    U_{\nu} = U_{\text{TBM}} U_{23} = U_{TM_1},
    \label{26}
\end{equation}
where $U_{\text{TBM}}$ denotes the TBM mixing matrix, and $U_{23}$ represents a unitary matrix that modifies the TBM structure, leading to the observed $\text{TM}_1$ mixing pattern. The diagonalization condition follows:  
\begin{equation}
    U_{\nu}^T m_\nu U_{\nu} = \text{diag}(m_1, m_2, m_3),
\end{equation}
where $m_1$, $m_2$, and $m_3$ correspond to the masses of the light neutrinos. The deviations from the exact TBM structure, particularly the nonzero value of the mixing angle $\theta_{13}$, are well accommodated within the $\text{TM}_1$ mixing scheme, making it consistent with neutrino oscillation data. Following provides the light neutrinos masses:

\begin{align}
    m_1 = 0,\quad
    m_2 = \frac{1}{2} \left| s - t + u \right|, \quad
    m_3 = \frac{1}{2} \left| s + t + u \right|,
\end{align}
where $s = 3a + 2b + 3c$,  $t = 24d^2$ and $u = (3a - 2b + 3c)^2$. The model suggests a normal hierarchy (NH) in neutrino masses, where $m_1 < m_2 < m_3$, as inferred from the above equation. The study of neutrino oscillation parameters, summarized in Table~\ref{tab:oscillation_parameters}, is crucial for understanding neutrino properties. These parameters include the mixing angles ($\theta_{12}$, $\theta_{23}$, $\theta_{13}$) and the mass-squared differences ($\Delta m^2_{21}$, $\Delta m^2_{31}$). They provide essential insights into neutrino mass and mixing behavior, describing the way neutrinos change flavor as they propagate. Future experimental programs, such as DUNE and Hyper-Kamiokande~\cite{DUNE:2020lwj,AliAjmi:2024xus}, aim to improve the precision of these parameters. Additionally, they will explore CP violation effects in the neutrino sector, offering deeper insight into fundamental symmetries of particle physics.

\section{Model Analysis and Predictions}
\label{sec:MA}
\begin{table}[t]
\centering
\renewcommand{\arraystretch}{1.2}
\setlength{\tabcolsep}{12pt}
\caption{The model parameters have been adjusted to align with the observed neutrino oscillation data~\cite{Esteban:2024eli}.}
\begin{tabular}{c c c} 
\hline
Parameter & Best-fit$\pm1\sigma$ & $3\sigma$ range \\ 
\hline
$\Delta m^2_{21}\ [10^{-5}\ \mathrm{eV}^2]$ & $7.49^{+0.19}_{-0.19}$ & $6.92 - 8.05$  \\  
$\Delta m^2_{31}\ [10^{-3}\ \mathrm{eV}^2]$ & $2.534^{+0.025}_{-0.023}$ & $2.463 - 2.606$  \\
$\sin^2 \theta_{12}$ & $0.307^{+0.012}_{-0.011}$ & $0.275 - 0.345$  \\  
$\sin^2 \theta_{23}$ & $0.561^{+0.012}_{-0.015}$ & $0.430 - 0.596$  \\  
$\sin^2 \theta_{13}$ & $0.02195^{+0.00054}_{-0.00058}$ & $0.02023 - 0.02376$  \\
\hline
\end{tabular}
\label{tab:oscillation_parameters}
\end{table}
\begin{figure*}[t]  
    \centering
    \begin{minipage}{0.48\textwidth}
        \includegraphics[width=\linewidth]{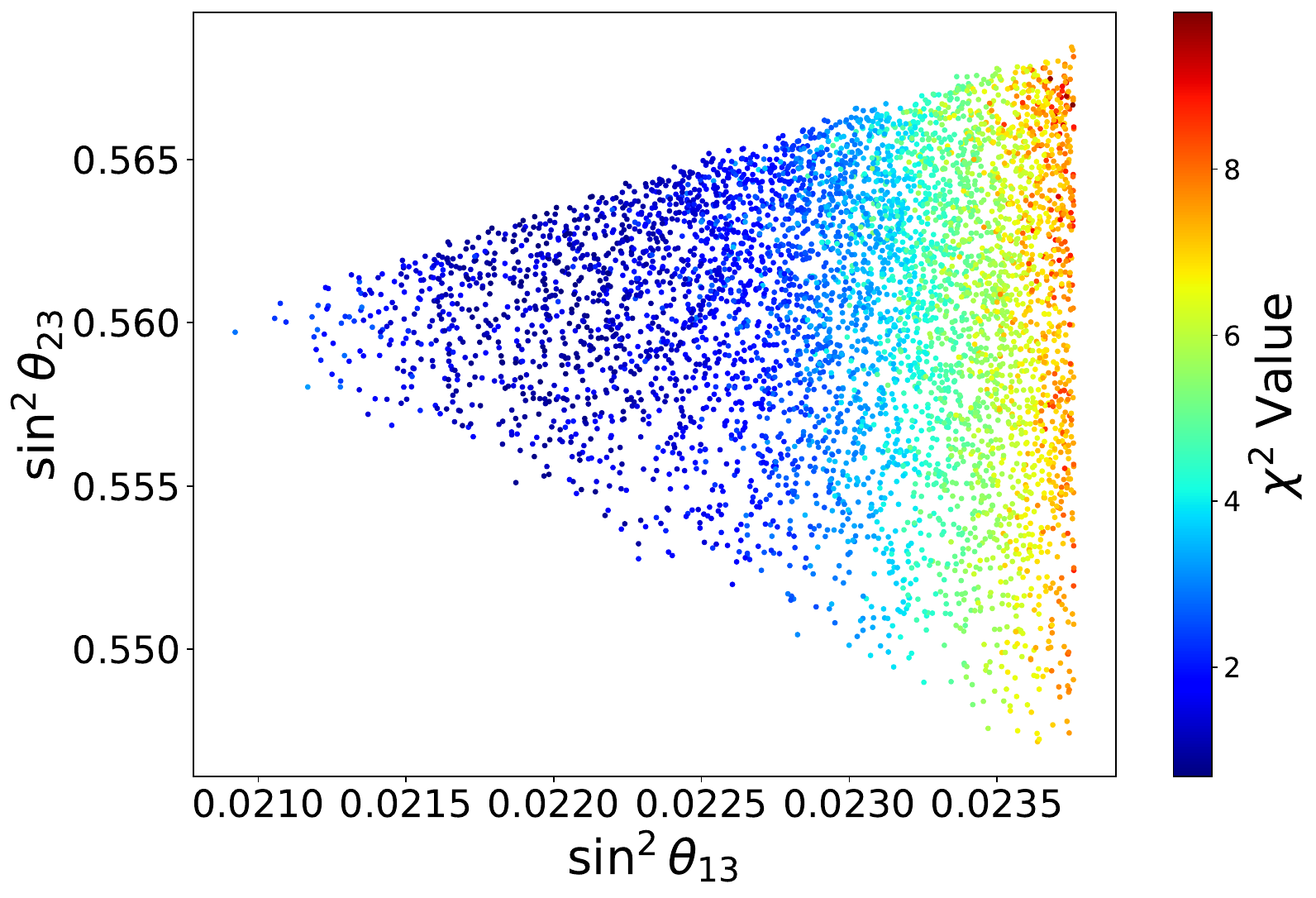}
        
    \end{minipage}\hfill
    \begin{minipage}{0.48\textwidth}
        \includegraphics[width=\linewidth]{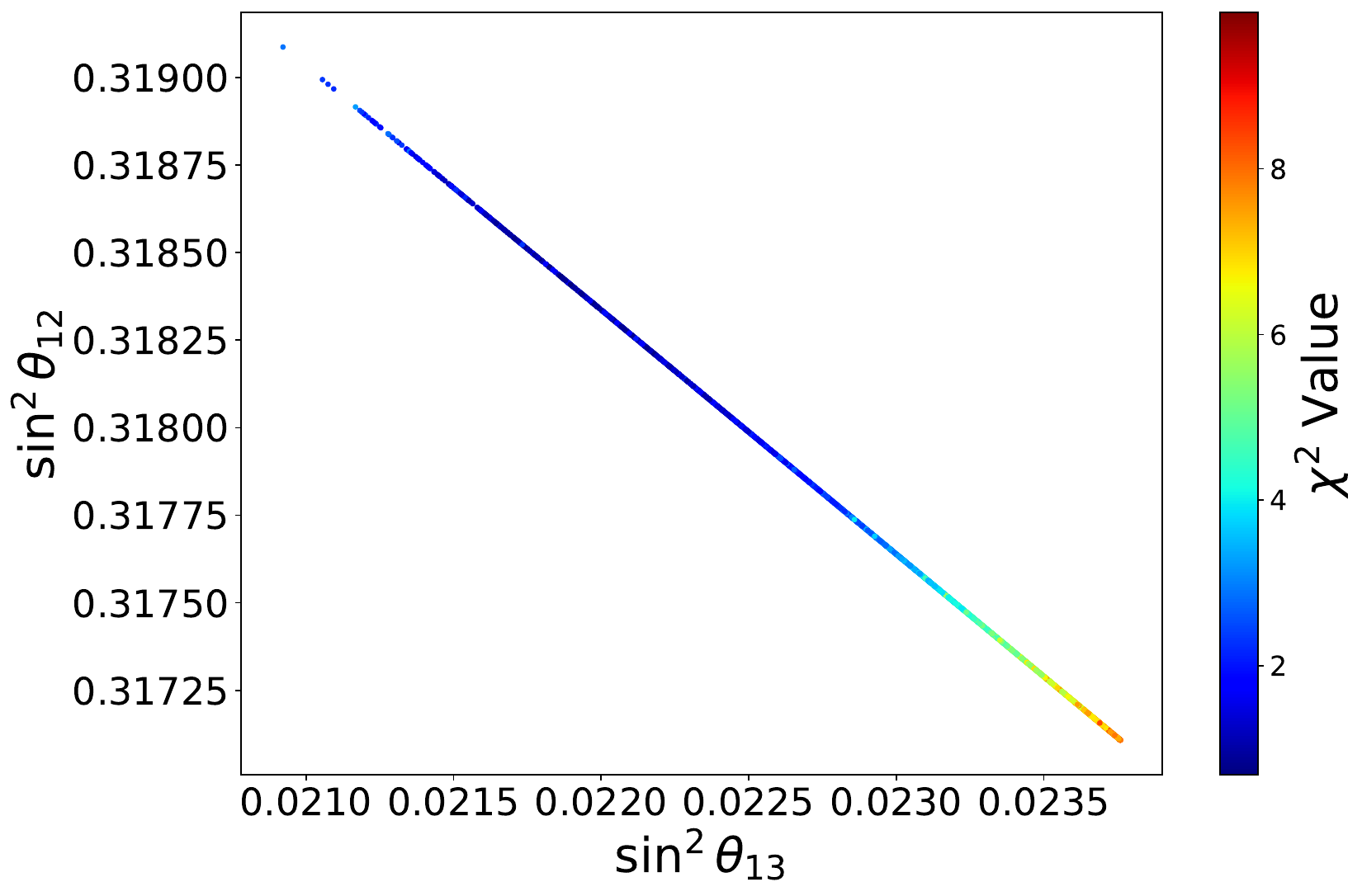}
        
    \end{minipage}
    
    \vspace{4mm} 

    \begin{minipage}{0.48\textwidth}
        \includegraphics[width=\linewidth]{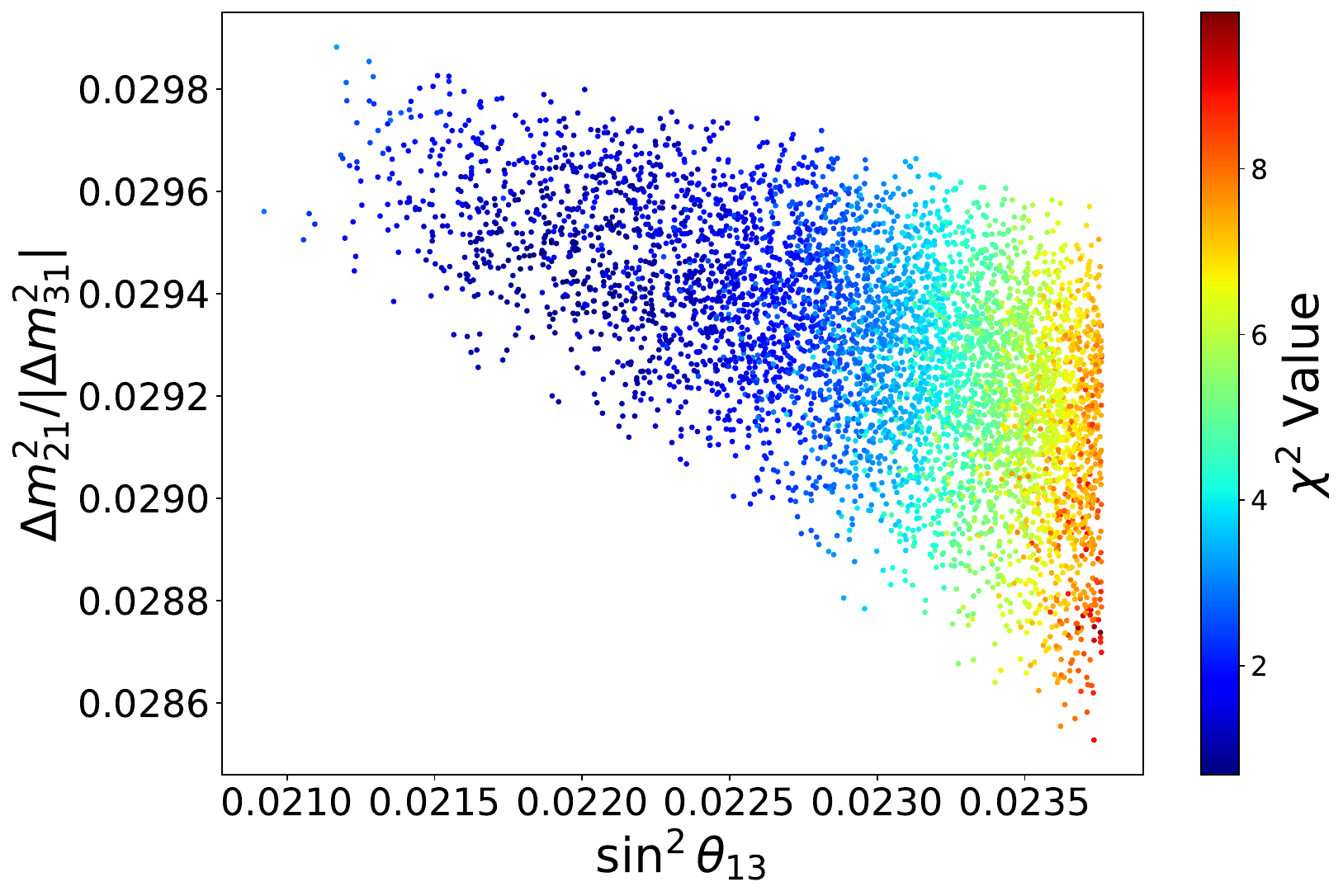}
        
    \end{minipage}\hfill
    \begin{minipage}{0.45\textwidth}
        \includegraphics[width=\linewidth]{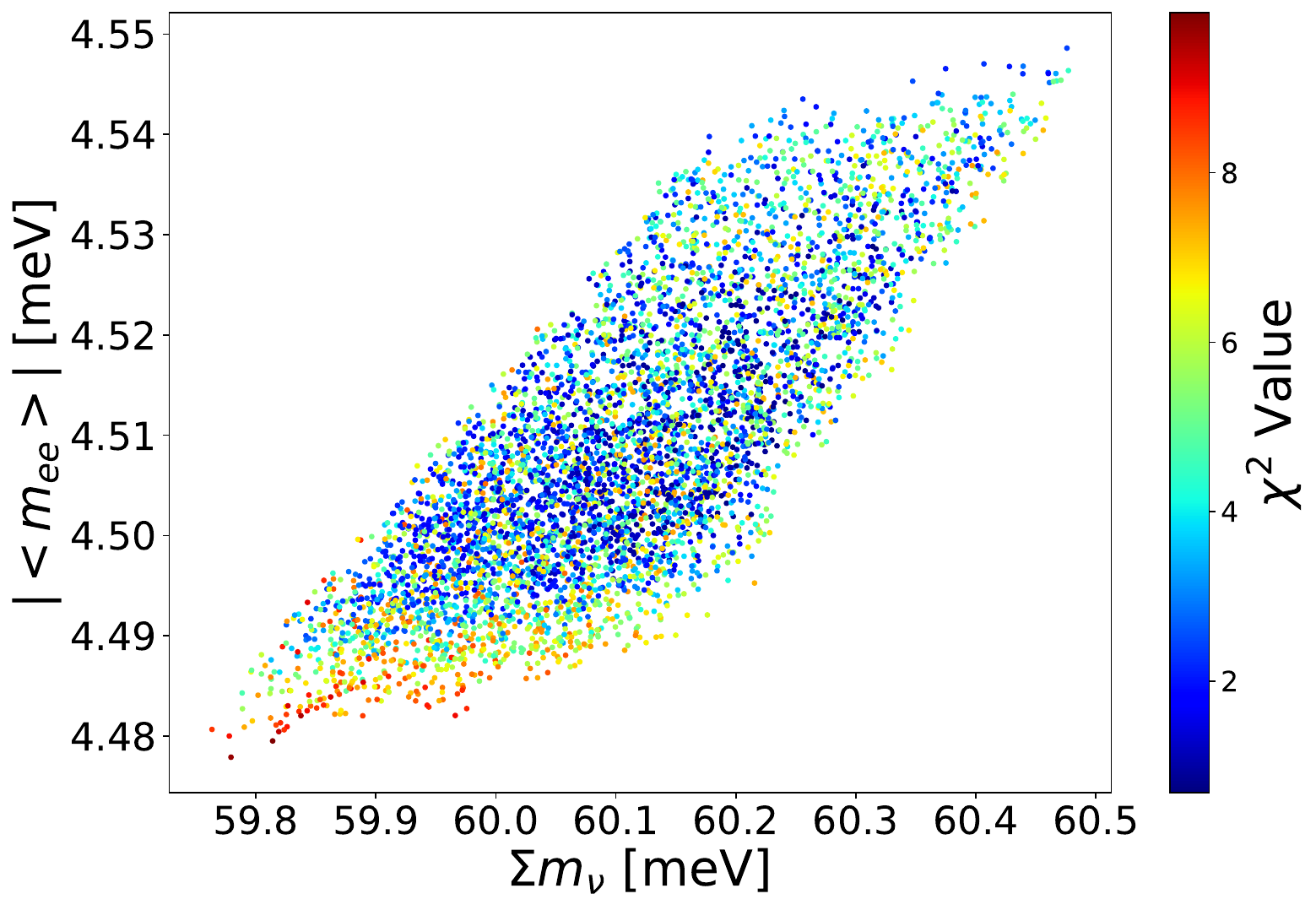}
        
    \end{minipage}

    \caption{ Relationship between neutrino oscillation parameters and the sum of the light neutrino masses with the effective Majorana electron neutrino mass.}
    \label{fig:1}
\end{figure*}
In the last section, we explored the $\nu$2HDM framework, which incorporates two RHNs and two scalar singlets under an $S_4 \otimes Z_4$ symmetry. Four real model parameters define the neutrino mass matrix in Eq.~\eqref{Eq:22}. A numerical analysis was performed to determine the values of these parameters in accordance with latest neutrino oscillation data~\cite{Esteban:2024eli}.

For the analysis, we adopted the 3$\sigma$ intervals for the neutrino oscillation parameters $\theta_{12}$, $\theta_{23}$, $\theta_{13}$, $\Delta m_{21}^2$ and $\Delta m_{31}^2$ as outlined in Table~\ref{tab:oscillation_parameters}. To focus on the solar and reactor neutrino parameters ($\Delta m_{21}^2$, $\theta_{12}$ and $\theta_{13}$) without the additional complexity introduced by atmospheric neutrino analyses, we used the C19 global fit, which excludes atmospheric neutrino data.
The parameters governing neutrino oscillations were reformulated in terms of the model’s fundamental quantities, while also incorporating the cosmological constraint on the sum of neutrino masses, which is restricted to \( \sum m_i< 0.12 \) eV as reported by Planck~\cite{Planck2018}.  

The model predicts the Dirac CP-violating phase in the range 
$\delta_{\mathrm{CP}} \in [73^\circ,\, 78^\circ]$, which lies outside the current $3\sigma$ experimental range. Future extensions incorporating additional CP-violating phases could resolve this tension.
The four real model parameters were treated as free variables and explored within the span:
\begin{align}
a \in [0.0108, 0.0109]\text{eV}, \quad 
b \in [0.0042, 0.0043]\text{eV}, \\
c \in [0.0063, 0.0064]\text{eV}, \quad
d \in [0.0009, 0.0016]\text{eV}.
\end{align}
The effective neutrino mass matrix (\( M_\nu \)), was numerically diagonalized using the following relation:
\begin{equation}
U^\dagger M_\nu U = \text{diag}(m_1^2, m_2^2, m_3^2), 
\end{equation} 
where the effective neutrino mass matrix is given by \( M_\nu = m_\nu m_\nu^\dagger \), and \( U \) represents a unitary matrix. The subsequent formula has been applied to get the mixing angles:
\begin{align}
&\sin^2 \theta_{13} = |U_{13}|^2,\quad
\sin^2 \theta_{23} = \frac{|U_{23}|^2}{1 - |U_{13}|^2} \\
& \qquad \quad \sin^2 \theta_{12} = \frac{1 - 3 \sin^2 \theta_{13}}{3 \cos^2 \theta_{13}} \label{Eq:31} \\
\cos \delta_{CP} & = \frac{(1 - 5 \sin^2 \theta_{13})(2 \sin^2 \theta_{23} - 1)}{4 \sin \theta_{13} \sin \theta_{23} \sqrt{2(1 - 3 \sin^2 \theta_{13})(1 - \sin^2 \theta_{23})}}
\end{align}

The correlations among the mixing angles observed in Fig.\ref{fig:1} arise naturally from the $\text{TM}_1$ mixing structure enforced by the $S_4 \otimes Z_4$ flavor symmetry. In this framework, the neutrino mixing angles $\theta_{12}$ and $\theta_{23}$ are not independent but are functionally related to $\theta_{13}$. Specifically, Eq.\eqref{Eq:31} explicitly links $\sin^2\theta_{12}$ to $\sin^2\theta_{13}$, while $\sin^2\theta_{23}$ is similarly constrained via the structure of the $\text{TM}_1$ matrix in Eq.~\eqref{26}. As $\theta_{13}$ increases, both $\theta_{12}$ and $\theta_{23}$ must adjust accordingly to preserve the $\text{TM}_1$ consistency, resulting in visible trends and correlations in the plots.

Furthermore, all viable parameter points derived from the $\chi^2$ minimization,
\begin{align}
  \chi^2 = \sum_i \frac{\left( \lambda_\text{model}^i - \lambda_\text{expt}^i \right)^2}{\Delta \lambda_i^2},  
\end{align}
where $\lambda_\text{model}^i$ and $\lambda_\text{expt}^i$ are the model predictions and experimental values for observable $i$, and $\sigma_i$ is the experimental $1\sigma$ uncertainty, fall within the $3\sigma$ bounds for $\theta_{12}$, $\theta_{23}$ and $\theta_{13}$. This demonstrates the model’s ability to simultaneously fit all mixing angles in a predictive, tightly constrained manner.
The following were determined to be the best-fit values for the parameters \(a \), \(b \), \(c \) and \(d \):
\begin{align}
 (a, b, c, d) = (0.01088, 0.00429, 0.00632, 0.00096) \text{eV}.   
\end{align}

The model parameters \( a \), \( b \), \( c \) and \( d \) are determined by the Yukawa couplings (\( Y_1, Y_2, Y_3 \)), the vev (\( \text{v}_2 \)) and the RHN masses (\( M_1, M_2 \)).
We consider  
\begin{equation}  
    M_1 \in [10^4, 10^6] \text{ GeV}, \quad M_2 \in [10^7, 10^8] \text{ GeV},  
\end{equation}  
with Yukawa couplings in the range \( \mathcal{O}(10^{-5}) \) to \( \mathcal{O}(10^{-2}) \). The chosen parameter ranges align well with the experimentally measured neutrino masses and mixing angles, ensuring theoretical consistency. The values which most closely fit our model are listed below:

\[
\sin^2 \theta_{12} = 0.318, \quad 
\sin^2 \theta_{23} = 0.558, \quad 
\sin^2 \theta_{13} = 0.0219,
\]
\[
\Delta m_{21}^2 = 7.46 \times 10^{-5} \, \text{eV}^2, \quad
\Delta m_{31}^2 = 2.53 \times 10^{-3} \, \text{eV}^2.
\]

This analysis confirms that the $\nu$2HDM with \( S_4 \otimes Z_4 \) symmetry can provide a viable explanation for neutrino oscillation data within the given constraints.

The effective Majorana mass \( \left| \langle m_{ee} \rangle \right| \), is a key parameter in neutrinoless double beta decay and is associated with the (1,1) element of the neutrino mass matrix when expressed in the flavor basis \cite{Thapa:2023fxu}. It is provided by the expression:

\begin{equation}
\left| \langle m_{ee} \rangle \right| = \left| \sum_{i=1}^{3} U_{ei}^2 \, m_i \right|,
\label{eq:mee_def}
\end{equation}

The parameters $U_{ei}$ correspond to the elements of the PMNS matrix linked to the electron flavor, while $m_i$ denote the neutrino mass eigenvalues. This term quantifies the combined contribution of neutrino mass eigenstates, each modulated by their mixing with the electron neutrino. In the case of a NH for neutrino masses, the effective Majorana mass is given by the following relation:

\begin{equation}
\left| \langle m_{ee} \rangle \right| \approx \left|  m_1U_{e1}^2 + m_2 U_{e2}^2 + m_3 U_{e3}^2 \right|,
\label{eq:mee_NH}
\end{equation}
where 
\begin{align}
U_{e1} = \cos\theta_{12} \cos\theta_{13},\ 
U_{e2} = \sin\theta_{12} \cos\theta_{13},\
U_{e3} = \sin\theta_{13} e^{-i\delta_{CP}}
\end{align}
Within the constraints of our model's parameter space, we calculated $\left| \langle m_{ee} \rangle \right|$, as shown in Fig.~\ref{fig:1}. The predicted values fall between 4 and 6 meV, which is well below the sensitivity limits of existing $0\nu\beta\beta$ experiments such as KamLAND-Zen, GERDA, and EXO-200. These experiments currently provide upper bounds on $\left| \langle m_{ee} \rangle \right|$ in the range of 20–60 meV ~\cite{KamLAND-Zen2016,GERDA2020,EXO-2002019}. Although current experiments are not sensitive to this small value, future ton-scale experiments like LEGEND-1000~\cite{collaboration2021legend1000} and nEXO~\cite{Adhikari_2021} may probe this regime.

\section{Bounds from Lepton flavor violating process}
\label{Sec:lFV}
LFV processes place stringent constraints on theoretical model parameters, particularly on the lepton number-violating Yukawa couplings (\( Y_{ij} \)). Among these processes, the rare decay \( \mu \to e\gamma \) serves as a sensitive probe for new physics. A stringent constraint on the branching ratio of this decay has been set by the MEG collaboration: $4.2 \times 10^{-13}$ ~\cite{MEG:2016}. The branching ratio of $\ell_{\alpha} \to \ell_{\beta}\gamma$ ~\cite{Guo:2017}:
\begin{equation}   
\text{Br}(\mu \to e \gamma) = \frac{3\alpha}{64\pi G^2} \left| \sum_i \frac{Y_{\mu i} Y_{ei}^*}{m_{H^+}^2} F\left(\Delta_{m_{H^+}}^{N_i}\right) \right|^2
\end{equation}
where $\Delta_{m_{H^+}}^{N_i} = (\frac{M_{N_i}}{m_{H^+}})^2$, where G is the Fermi constant and the expression for $F(x)$ is given by:
\begin{equation}
F(x) = \frac{1}{6(1-x)^4} \left[ 1 - 6x^2 + 3x^3 + 2x^3 - 6x^2 \ln x \right]
\end{equation}
\begin{figure}[t]
    \centering
    \includegraphics[width=0.45\textwidth]{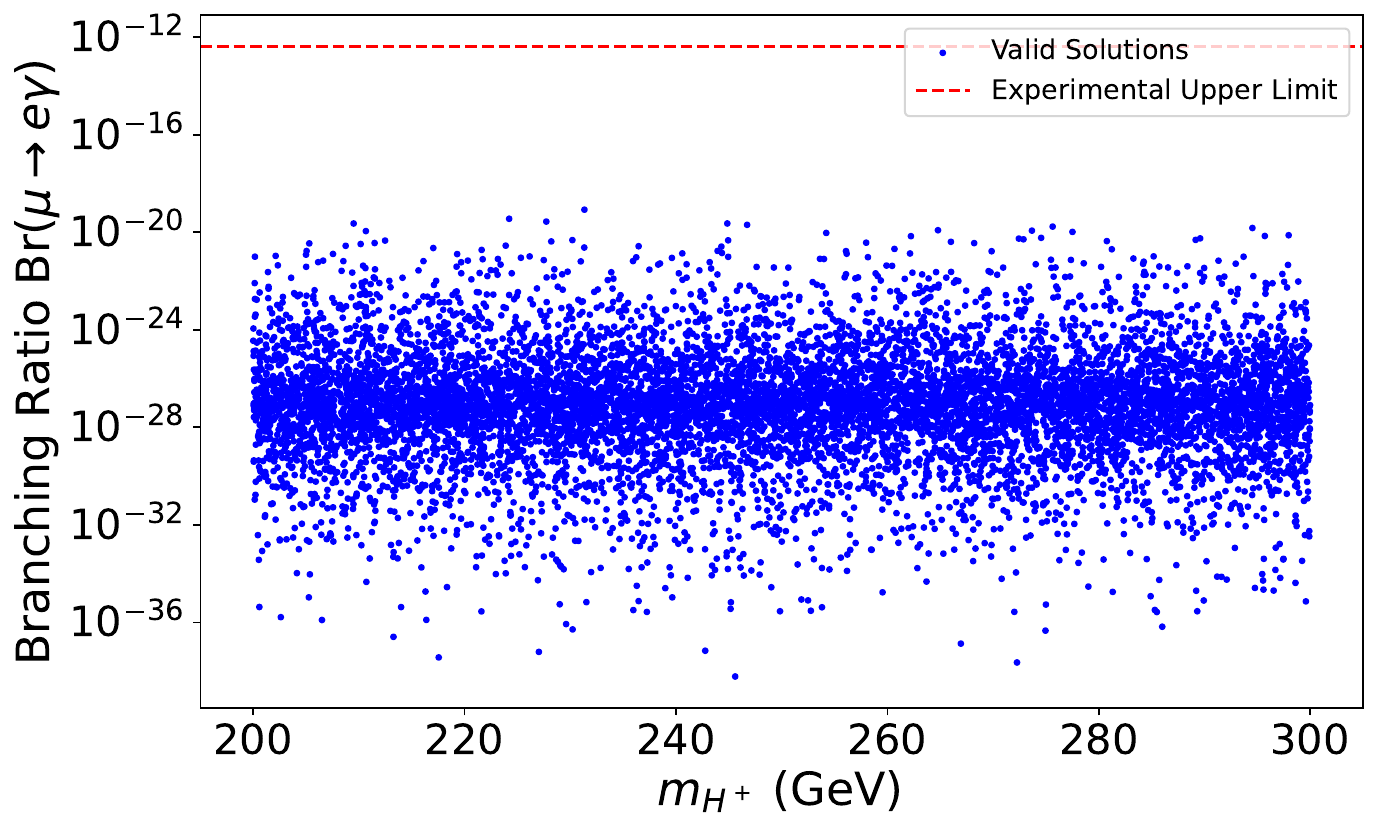}
    \caption{The branching ratio of $\mu \to e \gamma$ as a function of $m_{H^+}$. The dashed line indicates the experimental bound.}
    \label{fig:lfv}
\end{figure}

The model's Yukawa couplings are found to fall between $10^{-2}$ and $10^{-5}$, which is in line with the strict measured bounds on LFV processes. The expected branching ratio for \( \mu \to e\gamma \) is less than the experimental upper limit for NH of neutrino masses, as illustrate in Figure~\ref{fig:lfv}.

These findings validate the consistency of the Yukawa couplings within the model and highlight their compliance with experimental constraints. Furthermore, the model remains robust in addressing broader phenomena such as leptogenesis and dark matter, emphasizing its potential to explain critical aspects of particle physics.

\section{Leptogenesis}
\label{Sec:lept}

Observations indicate that the Universe contains more matter than antimatter, a phenomenon known as BAU. To explain this asymmetry, a theoretical mechanism called leptogenesis has been proposed. Leptogenesis begins by producing a lepton imbalance via RHN decays. These decays result in a small excess of leptons over antileptons because of CP violation. The observed asymmetry between matter and antimatter in the Universe is linked to sphaleron-induced processes, which convert some portion of the lepton asymmetry into baryonic matter~\cite{Rubakov:1996}.

In standard leptogenesis scenarios, successful baryogenesis typically requires the lightest RHN (\(N_1\)) to have a mass exceeding \(10^9\) GeV \cite{Davidson:2002,Buchmuller:2002}. However, in this study, we explore an alternative framework where \(M_1\) is significantly lower as \(10\) TeV. By optimizing CP asymmetry and carefully choosing model parameters, leptogenesis remains feasible even at these lower energy scales. Our model assumes a hierarchical mass structure, where the Higgs doublet mass \(m_{\phi_2}\) is much smaller than \(M_1\), which in turn is considerably smaller than \(M_2\), the next heaviest RHN mass.

The parameter responsible for CP violation, denoted as (\(\epsilon_1\)), is a key element in this framework. It arises from the decay of \(N_1\) and is mathematically represented as~\cite{Borah2019}:

\begin{equation}
    \epsilon_1 = \frac{1}{8\pi (Y^\dagger Y)_{11}} \sum_{j \neq 1} \text{Im}[(Y^\dagger Y)^2]_{1j} \left[ \mathcal{F}(r_{j1}, \eta_1) - \sqrt{\frac{r_{j1}}{r_{j1} - 1}} (1-\eta_1)^2 \right],
\end{equation}
where \( \mathcal{F}(r_{j1}, \eta_1) \) is given by:
\begin{equation}
    \mathcal{F}(r_{j1}, \eta_1) = \sqrt{r_{j1}} \left[ 1 + \frac{1 - 2\eta_1 + r_{j1}}{(1-\eta_1)^2} \ln\left(\frac{r_{j1} - \eta_1^2}{1 - 2\eta_1 + r_{j1}}\right) \right].
\end{equation}
Here, mass ratios in the system are described by \( r_{j1} = \left(\frac{M_{j}}{M_1}\right)^2 \) and \( \eta_1 = \left(\frac{m_{\phi_2}}{M_1}\right)^2 \).
Eq.~\eqref{eq:20} parameterizes the Yukawa coupling matrix (\( Y \)), which describes the interaction between massive RHN's, light neutrinos and the Higgs field.
 The result is as follows:
\begin{equation}
    Y = \frac{\sqrt{2}}{\text{v}_2} U_{\text{PMNS}} (m_\nu^{\text{diag}})^{1/2} R (M_N^{\text{diag}})^{1/2},
    \label{eq:casa}
\end{equation}
leading to the relation:
\begin{equation}
    Y^\dagger Y = \frac{2}{\text{v}_2^2} (M_N^{\text{diag}})^{1/2} R^\dagger m_\nu^{\text{diag}} R (M_N^{\text{diag}})^{1/2}.
\end{equation}
We examine two RHNs in our scenario, where the heaviest one either decouples because of its large mass or has a low Yukawa coupling. Light neutrino masses are denoted by \( m_1 \), \( m_2 \) and \( m_3 \). For NH, \( m_1 \to 0 \) and \( R \) are orthogonal matrices parameterized by a complex angle \( \theta \)~\cite{Antusch:2011nz}:
\begin{equation}
R =
\begin{bmatrix}
\cos \theta & \sin \theta \\
-\sin \theta & \cos \theta
\end{bmatrix}.
\end{equation}
Given $N_1$, the decay width ($\Gamma_1$) that establishes the rate of decay into leptons and Higgs bosons is as follows:
\begin{equation}
    \Gamma_1 = \frac{M_1}{8\pi} (Y^\dagger Y)_{11} (1-\eta_1)^2,
\end{equation}
Here, $M_1$ is the mass of $N_1$, $(Y^\dagger Y)_{11}$ represents the effective coupling strength of $N_1$ to the leptons and Higgs field.
The decay parameter associated with \(N_1\), denoted as \(K_{N_1}\) is given by:
\begin{equation}
    K_{N_1} = \frac{\Gamma_1}{H(z=1)},
\end{equation}
Here \(z = M_1/T\) and Hubble parameter is denoted by \( H \), where T represents the temperature of the thermal bath.
\begin{equation}
    H = \sqrt{\frac{8\pi^3 g_*}{90}} \frac{T^2}{M_{\text{Pl}}},
\end{equation}
The parameter \( M_{\text{Pl}} \) refers to the fundamental mass scale associated with gravity, while \( g_* \) quantifies the count of relativistic degree of freedom in the early Universe. The efficiency of leptogenesis is largely determined by the decay parameter \( K_{N_1} \), which describes whether the decays of \( N_1 \) take place in thermal equilibrium (\( K_{N_1} \gg 1 \)) or out of equilibrium (\( K_{N_1} \ll 1 \)).
\begin{equation}
    K_{N_1} \approx 897 \left({\tan \beta}\right)^2 \frac{|((m_\nu^{\text{diag}}) ^R)_{11}|}{\text{eV}},
\end{equation}
where \( (m_\nu^{\text{diag}})^R = R^\dagger m_\nu^{\text{diag}} R \), which simplifies to:
\begin{equation}
    ((m_\nu^{\text{diag}})^R)_{11} =  m_2\cos^2 \theta + m_3 \sin^2 \theta.
\end{equation}

To explore the parameter space and study the viability of leptogenesis, we vary $\theta$ over a random range, enabling us to examine the dependency of $K_{N_1}$ and the resulting baryon asymmetry on this key parameter. This framework enhances our understanding of the conditions necessary for successful leptogenesis, particularly within low-scale models. The dynamics of the number density of the lightest RHN (\( N_1 \)) and the resulting \( B-L \) asymmetry evolve according to a set of coupling Boltzmann equations~\cite{Davidson:2002}.

\begin{align}  
    \frac{d n_{N_1}}{dz} &= -D_1 (n_{N_1} - n_{N_1}^{\text{eq}}),  \\  
    \frac{d n_{B-L}}{dz} &= -\epsilon_1 D_1 (n_{N_1} - n_{N_1}^{\text{eq}}) - W_1 n_{B-L}.
\end{align}  
Here, \( n_{N_1} \) represents the number density of \( N_1 \), while \( n_{B-L} \) denotes the net lepton asymmetry generated by its decay. The equilibrium distribution for the number density of \( N_1 \) is expressed as:

\begin{equation}  
    n_{N_1}^{\text{eq}} = \frac{z^2}{2} K_1(z),  
\end{equation}  
where \( K_1(z) \) is the second-kind modified Bessel function.

The decay term \( D_1 \) determines the efficiency of \( N_1 \) decays and is defined as  
\begin{equation}  
    D_1 \equiv \frac{\Gamma_1}{H z} = \frac{K_{N_1} z}{K_2(z)},  
\end{equation}  
where \( K_2(z) \) is another modified Bessel function. The reduction in the imbalance of \( B-L \) is explained by the total washout term, \( W_1 \), which is given by:  
\begin{equation}  
    W_1 = W_{1D} + W_{\Delta L=2}.  
\end{equation}  
\begin{figure*}[h]  
    \centering
    \begin{minipage}{0.45\textwidth}
        \includegraphics[width=\linewidth]{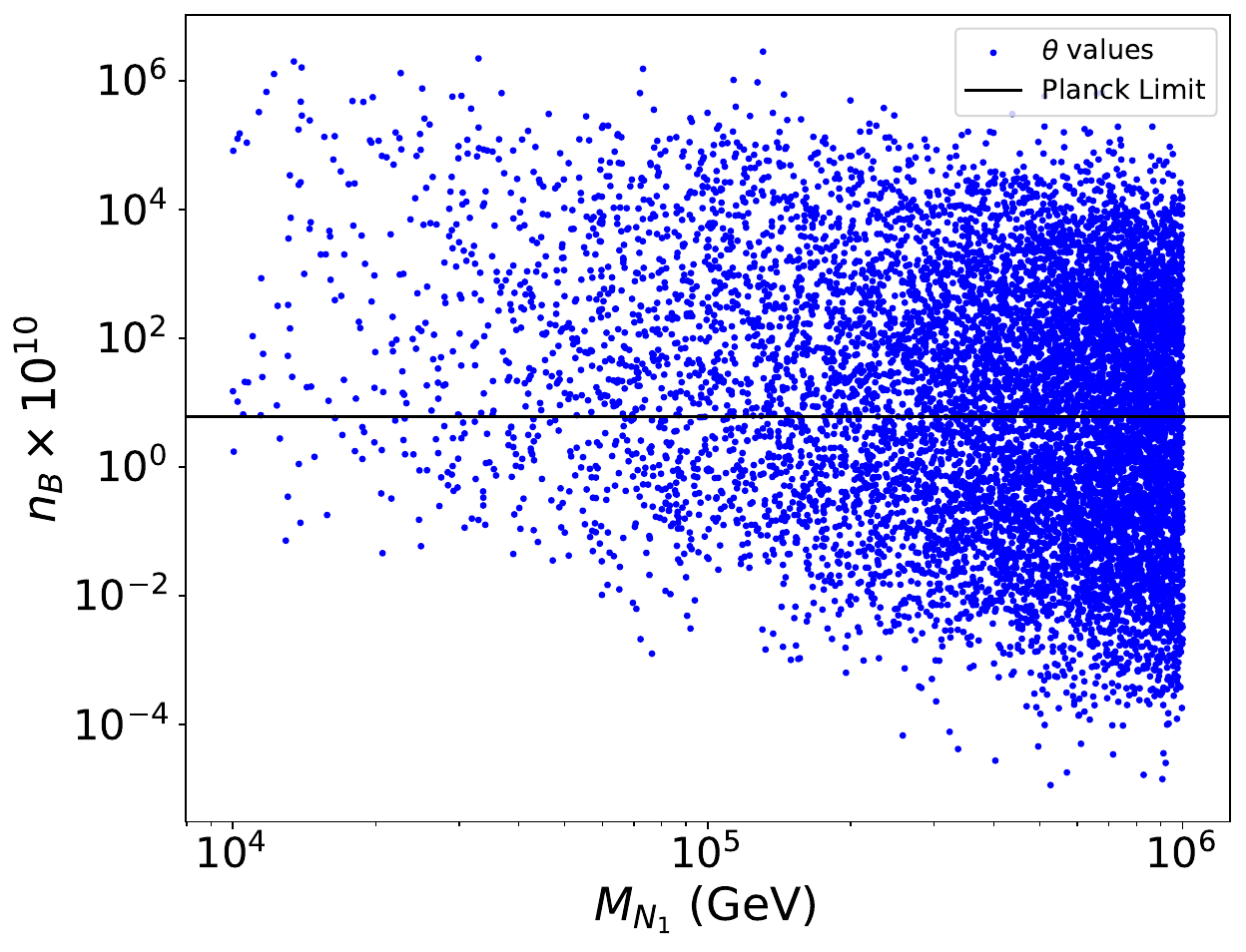}
        
    \end{minipage}\hfill
    \begin{minipage}{0.45\textwidth}
        \includegraphics[width=\linewidth]{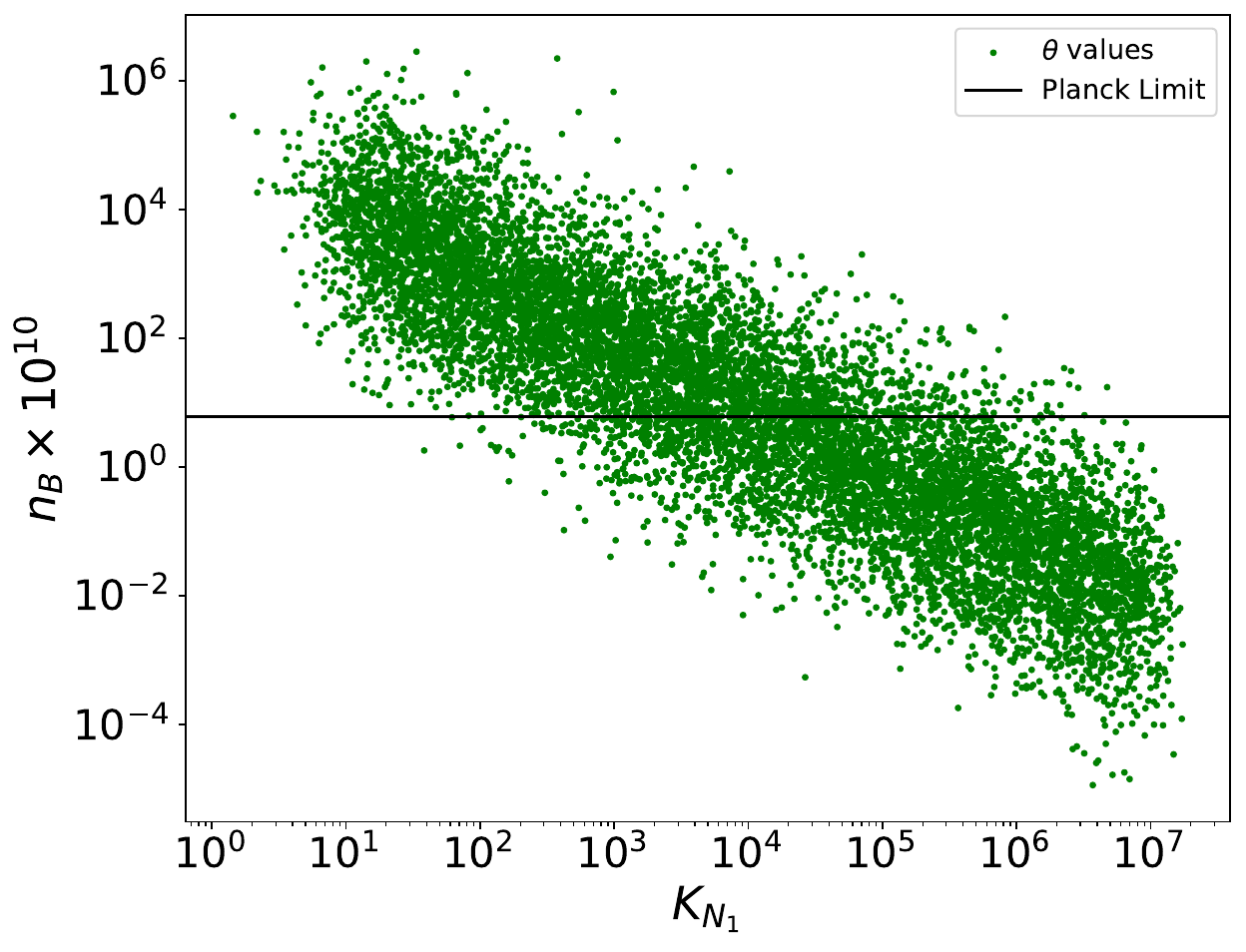}
        
    \end{minipage}
    
    \vspace{4mm} 

    \begin{minipage}{0.45\textwidth}
        \includegraphics[width=\linewidth]{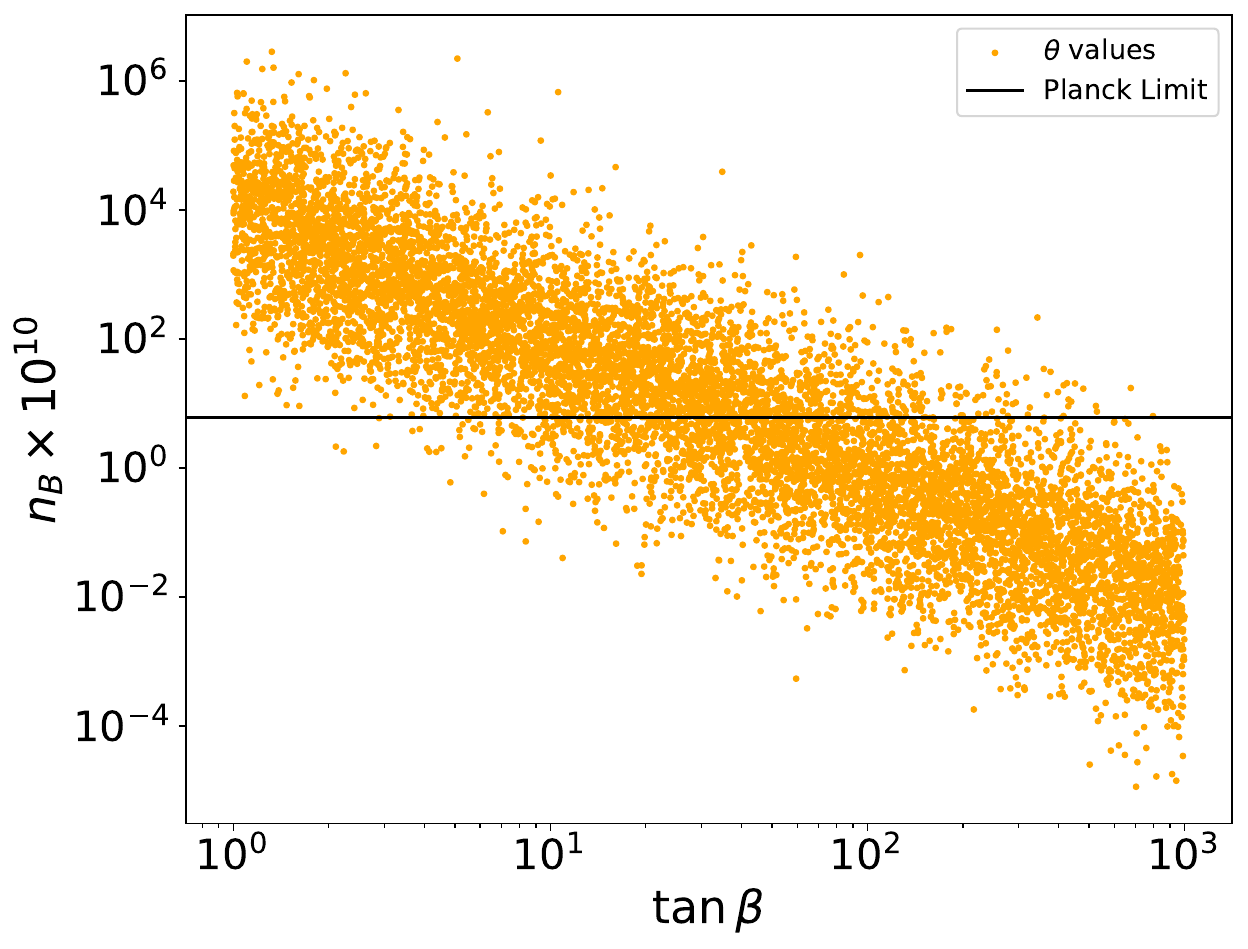}
        
    \end{minipage}\hfill
    \begin{minipage}{0.45\textwidth}
        \includegraphics[width=\linewidth]{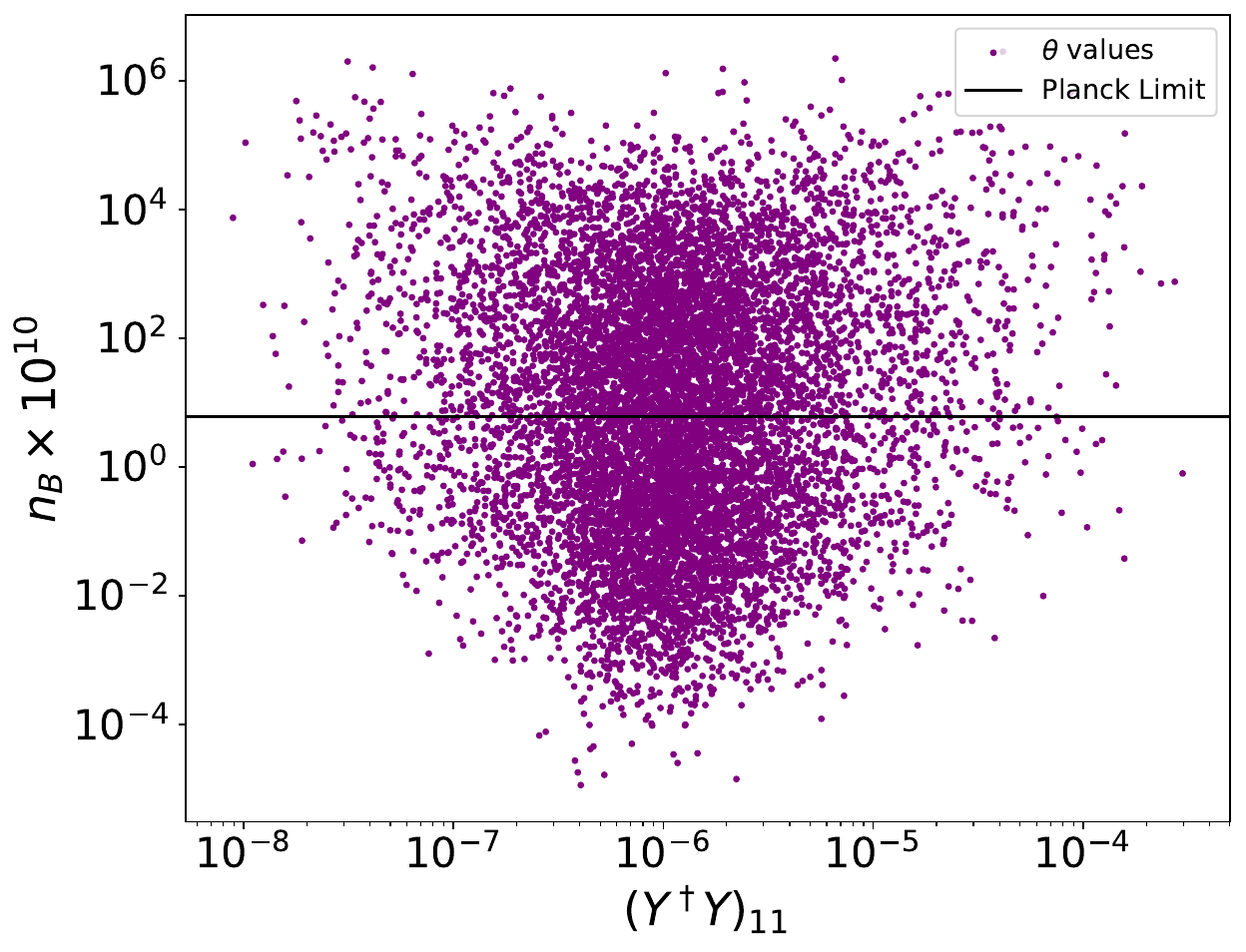}
        
    \end{minipage}

    \caption{The universe's baryon asymmetry is shown here using the random scan over angle $\theta$ for NH expressed in terms of $M_{N_1}$, $K_{N1}$, $tan\beta$ and $(Y^\dagger Y)_{11}$, respectively.}
    \label{fig:all_plots}
\end{figure*}
\begin{figure*}[!t]
    \centering
    \includegraphics[width=\linewidth]{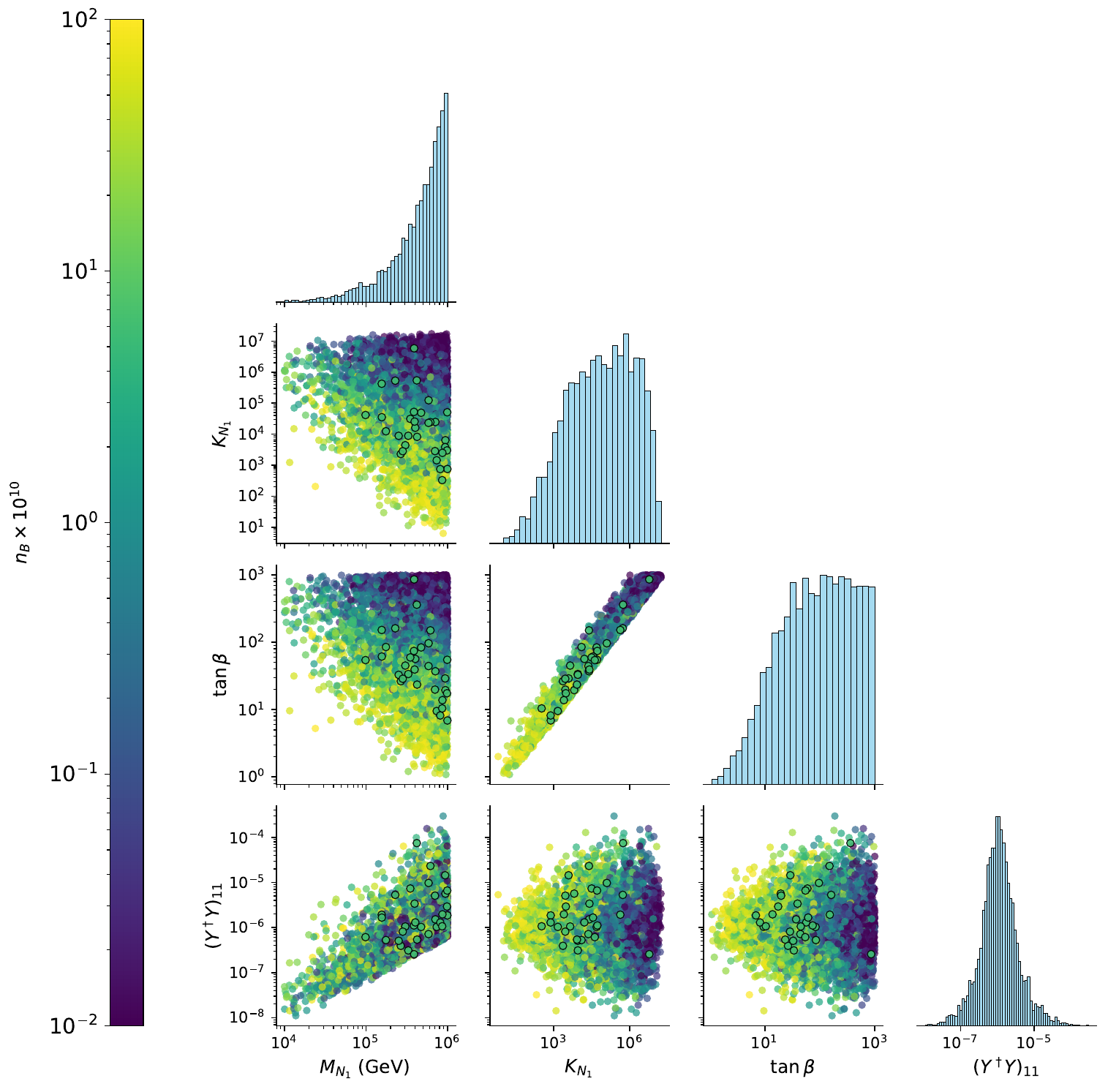}
    \caption{Lower-triangular correlation plot of $M_{N_1}$, $K_{N1}$, $\tan\beta$ and 
$(Y^\dagger Y)_{11}$ for the NH scenario. Color indicates the predicted $n_{B} \times 10^{10}$, 
and black circles mark points consistent with the Planck BAU measurement.}

    \label{fig:corelation}
\end{figure*}
The inverse decay contribution, arising from processes such as \( l_L \phi_2, \bar{l}_L \phi_2^* \to N_1 \), is expressed as  
\begin{equation}  
    W_{1D} = \frac{1}{4} K_{N_1} z^3 K_1(z).  
\end{equation}  
Furthermore, $\Delta{L}=2$ lepton-number-violating scatterings (\( l_L \phi_2 \leftrightarrow \bar{l}_L \phi_2^* \), \( l_Ll_L \leftrightarrow \phi_2^* \phi_2^* \)) cause the washout, which is generally represented as~\cite{Borah2019}:  
\begin{equation}  
    W_{\Delta L=2} \simeq  {\frac{18}{\pi^4}} \frac{\sqrt{10} M_{\text{Pl}}}{g_l \sqrt{g_*} z^2 \text{v}_2^4}  \left( \frac{2\pi^2}{\lambda^5} \right)^2 M_1 \bar{m}^2.  
\end{equation}  
In this context, \( g_l \) denotes the intrinsic degrees of freedom associated with the SM leptons, while \( \bar{m} \) characterizes the relevant neutrino mass scale, given by:
  
\begin{equation}  
    \bar{m}^2 = m_1^2 + m_2^2 + m_3^2.  
\end{equation}  

The sphaleron conversion factor that connects the ultimate baryon asymmetry $n_B$ to the resulting $B-L $ asymmetry can be expressed as:

\begin{equation}  
    n_B \approx \frac{3}{4} \frac{g_0^*}{g_*} a_{\text{sph}} n_{B-L}^{\text{final}},  
\end{equation}  
During the initial stages of the Universe, the number of relativistic degrees of freedom are characterized by \( a_{\text{sph}} = \frac{8}{23} \), \( g_0^* = \frac{43}{11} \), and \( g_* = 110.75 \). The Planck 2018 collaboration provides an estimate for the observed baryon asymmetry~\cite{Planck2018}.

\begin{equation}  
    n_B = (6.04 \pm 0.08) \times 10^{-10}.  
\end{equation}

\section{Numerical Analysis}
\label{Sec:NA}

The previous section~\ref{Sec:lept} explores the viability of leptogenesis at the TeV scale, considering a scenario with two RHNs. The discussion particularly focuses on the role of Yukawa interactions, RHN mass spectrum and CP-violating phases in this mechanism.  A crucial component of our study is making sure that the Yukawa couplings are successful in BAU while staying within the limits imposed by lepton flavor violation requirements.

We use the Casas-Ibarra parameterization in Eq.~\eqref{eq:casa} to describe the neutrino sector. This method helps systematically explore different values of Yukawa couplings and RHN masses. In Section~\ref{sec:MA}, we showed that the RHN masses are chosen within specific ranges,  $M_1=10^4 - 10^6 \text{ GeV}$ and $M_2 = 10^7 - 10^8 \text{ GeV}$ to ensure consistency with the Yukawa couplings and the vev of $\phi_2$. This hierarchical structure is crucial for generating the required CP asymmetry and ensuring that leptogenesis occurs at the TeV scale.
The rotational matrix \(R\), which parameterizes the RHN contributions, introduces an arbitrary complex angle \(\theta\). This angle is crucial in establishing the CP asymmetry and the Universe's baryon asymmetry. For our analysis, \(\theta\) is varied across a wide range from \((10^{-2} + i \cdot 10^{-2})\) to \((0.5 \times 10^{-1} + i \cdot 0.5 \times 10^{-1})\). This variation allows us to comprehensively study the dependency of leptogenesis parameters on \(\theta\).

In the NH scenario, Fig.~\ref{fig:all_plots} illustrates how the BAU depends on important parameters, such as the RHN mass \( M_1 \), the decay parameter \( K_{N_1} \), \(\tan \beta\) and the Yukawa couplings. Our findings demonstrate that successful baryogenesis can be achieved across a broad range of \( M_1 \) values, with the most promising region lying between \( 10^5 - 10^6 \) GeV. A considerable number of parameter points in this mass range satisfy the Planck limits, supporting the feasibility of TeV-scale leptogenesis.  

The decay parameter \( K_{N_1} \) serves as a crucial measure of how efficiently the RHN decays occur relative to the Universe’s expansion rate. At low $K_{N1}$, the RHN decays are much slower than the Hubble expansion rate, so the washout from inverse decays is weak. This allows most of the generated lepton asymmetry to survive, giving a large BAU. As $K_{N1}$ increases, the decays become faster and the washout processes strengthen, erasing more of the asymmetry. From the \( K_{N_1} \) vs BAU plot, we observe that parameter points satisfying the Planck constraints predominantly fall within the range \( K_{N_1} \sim 10^{3} - 10^{6} \), indicating the optimal balance required for successful leptogenesis. This range underscores the delicate interplay between decay rates and equilibrium dynamics in baryogenesis.  

The parameter \( \tan \beta \), which governs the Higgs sector interactions, plays a significant role in BAU generation. For smaller values of $\tan\beta$, the washout from Higgs-mediated processes is weaker, allowing a larger baryon asymmetry to survive. As $\tan\beta$ increases, these washout effects become stronger, suppressing the generated asymmetry and leading to a monotonic decrease of BAU across the scanned range. From the \( \tan \beta \) vs BAU plot, we find that values between 10 and 100 are consistent with successful baryogenesis, contingent on variations in the mixing angle \( \theta \). This highlights the non-trivial impact of Higgs dynamics on leptogenesis.  

The Yukawa couplings primarily govern the CP asymmetry and the decay rates of RHNs. The plot of \( Y^\dagger Y \) vs BAU reveals a critical balance in this parameter space. When the Yukawa couplings are very small, the CP asymmetry generated is insufficient to produce the required BAU. On the other hand, if they are too large, the system reaches equilibrium too quickly, erasing any generated asymmetry through strong washout effects. This results in a narrow range of Yukawa coupling values, typically between \( 10^{-7} - 10^{-5} \), where successful baryogenesis can occur while maintaining consistency with Planck constraints.

While the single-parameter projections in Figure~\ref{fig:all_plots} clearly show the 
individual dependence of BAU on each variable, the lower-triangular correlation plot 
(Figure~\ref{fig:corelation}) additionally reveals how pairs of parameters are correlated 
and which specific combinations lead to BAU values consistent with the Planck measurement. 
The diagonal panels display the distribution of each scanned parameter, while the 
off-diagonal panels illustrate the variation of the baryon asymmetry with parameter pairs, 
with the color scale indicating the predicted $n_{B} \times 10^{10}$. Black circles denote 
points satisfying the Planck measurement of the BAU. This multidimensional view complements 
the single-parameter plots and provides a direct visualization of the combined parameter 
space consistent with the observed BAU.

Assuming suitable RHN masses, mixing angles and limited ranges of \( K_{N_1} \), \( \tan \beta \) and the Yukawa couplings, our numerical study demonstrates that TeV-scale leptogenesis is possible inside the NH framework. The findings highlight how important it is to keep these parameters in a careful balance in order to promote successful baryogenesis and prevent excessive washout effects.

\section{Conclusion}
\label{Sec:conclusion}
The $\nu$2HDM with $S_4 \otimes Z_4$ flavor symmetry, which inevitably results in the $\text{TM}_1$ structure of mixing in the leptonic sector, was studied in this work. This work investigates the theoretical predictions for neutrino masses, considering a mass structure where the lightest neutrino has the smallest mass. The estimated mass values fall within the range of \(0.0597\) eV to \(0.0605\) eV. Apart from the CP phase, which remains unconstrained, we executed a numerical exploration of the parameter space of the model and identified a region within a confidence level \(3\sigma\) where the predicted mixing angles and the mass-squared differences align with the experimental observations. To refine these predictions, a chi-squared analysis was conducted, allowing us to determine the optimal parameter values. Additionally, based on the obtained parameters, we calculated the effective Majorana neutrino mass \( |\langle m_{ee} \rangle| \), finding it to be extremely small. This suggests that current \(0\nu\beta\beta\) experiments may face significant challenges in detecting it.
Additionally, we validate the model by testing its predictions against LFV constraints. The Yukawa couplings predicted in our model remain consistent with low-energy experimental constraints, particularly the  highest allowed branching ratio for \( \mu \to e\gamma \) is restricted by experimental observations, which is set at \( 4.2 \times 10^{-13} \) by the MEG Collaboration~\cite{MEG:2016}.\\
Additionally, we conducted a comparative examination of leptogenesis by examining the effects of changes in the rotational matrix's arbitrary angle on our findings. The Yukawa coupling matrix, which is important for many phenomenological characteristics of the model, has been determined numerically. By selecting an appropriate rotational matrix $R$ and adjusting the arbitrary angle, we studied its impact on cosmological phenomena. Inspired by earlier studies \cite{Hugle:2018qbw,2016}, which explored TeV-scale leptogenesis with specific choices of RHN masses and the vev of $\phi_2$, we propose modified values tailored to our framework. We then verify the consistency of these parameters with the Yukawa couplings, ensuring compatibility with neutrino phenomenology. We considered the variation of $M_{1}$ in the range $10^4 - 10^6$ GeV and $M_{2}$ within $10^7 - 10^8$ GeV, allowing us to plot relevant parameters against the BAU, with the observed BAU represented as a reference line.    
Our numerical results highlight key leptogenesis parameters, including $K_{N_1}$, $M_1$, $\tan \beta$ and \( Y^\dagger Y \). We demonstrated that TeV-scale leptogenesis is achievable within the NH framework when RHN masses, $\theta$ values, and constrained parameter ranges are appropriately chosen. The measured BAU can be consistently explained by the interplay of these parameters, reinforcing the viability of low-scale leptogenesis.

Overall, our study demonstrates that the proposed model successfully explains neutrino observables and leptogenesis while remaining consistent with experimental constraints. These results strongly support the observed BAU and provide a solid basis for leptogenesis at the TeV scale.
\section*{Acknowledgment}  
The corresponding author acknowledges the Ministry of Human Resource Development (MHRD), Government of India, for the financial support provided through the GATE Senior Research Fellowship (SRF).

\bibliographystyle{model5-num-names}

\begin{thebibliography}{10}

\bibitem{deSalas:2017kay}
P.~F. de Salas, D.~V. Forero, C.~A. Ternes, M.~Tortola and J.~W.~F. Valle,
\newblock Status of neutrino oscillations 2018: $3\sigma$ hint for normal mass ordering and improved CP sensitivity,
\newblock {\em Phys. Lett. B} {782}, 633--640 (2018),
\newblock arXiv:1708.01186 [hep-ph].

\bibitem{PDG2020}
Particle~Data Group, P.~A. Zyla, R.~M. Barnett, J.~Beringer, O.~Dahl, D.~A. Dwyer, D.~E. Groom, C.-J. Lin, K.~S. Lugovsky, E.~Pianori, and et~al.
\newblock Review of Particle Physics,
\newblock {\em PTEP} { 2020}, no. 8, 083C01 (2020).

\bibitem{buchmuller2005}
W.~Buchm\"uller, R.~D.~Peccei and T.~Yanagida,
\newblock Leptogenesis as the origin of matter,
\newblock {\em Annu. Rev. Nucl. Part. Sci.} {55}, 311--355 (2005), arXiv:hep-ph/0502169.

\bibitem{ma2006}
E.~Ma,
\newblock Verifiable radiative seesaw mechanism of neutrino mass and dark matter,
\newblock {\em Phys. Rev. D} {73}, 077301 (2006), arXiv:hep-ph/0601225
\newblock [hep-ph/0601225].

\bibitem{fukuda1998}
Y.~Fukuda et~al. (Super-Kamiokande~Collaboration).
\newblock Evidence for oscillation of atmospheric neutrinos.
\newblock {\em Phys. Rev. Lett.} {81},1562--1567 (1998), arXiv:hep-ex/9807003.

\bibitem{ahmad2001}
Q.~R.~Ahmad et~al. (SNO~Collaboration).
\newblock Measurement of the rate of $\nu_e + d \rightarrow p + p + e^-$ interactions produced by $^8$b solar neutrinos at the sudbury neutrino observatory.
\newblock {\em Phys. Rev. Lett.} {87}, 071301 (2001), arXiv:nucl-ex/0106015.

\bibitem{minkowski1977}
P.~Minkowski,
\newblock $\mu \to e\gamma$ at a rate of one out of $10^{9}$ muon decays?,
\newblock {\em Phys. Lett. B} {67}, 421--428 (1977).

\bibitem{Mohapatra1980}
R.~N.~Mohapatra and G.~Senjanovi\'{c},
\newblock Neutrino mass and spontaneous parity nonconservation,
\newblock {\em Phys. Rev. Lett.} {44}, 912--915 (1980).

\bibitem{Vissani1998}
F.~Vissani,
\newblock Do experiments suggest a hierarchy problem?,
\newblock {\em Phys. Rev. D} {57}, 7027--7030 (1998), arXiv:hep-ph/9709409.

\bibitem{Pilaftsis2004}
A.~Pilaftsis and T.~E.~J.~Underwood,
\newblock Resonant leptogenesis,
\newblock {\em Nucl. Phys. B} {692}, 303--345 (2004), arXiv:hep-ph/0309342.

\bibitem{akhmedov1998}
E.~Kh. Akhmedov, V.~A. Rubakov, and A.~Yu. Smirnov.
\newblock Baryogenesis via neutrino oscillations.
\newblock {\em Phys. Rev. Lett.} {81}, 1359--1362 (1998), arXiv:hep-ph/9803255.

\bibitem{Asaka2005}
T.~Asaka and M.~Shaposhnikov.
\newblock The $\nu$msm, dark matter and baryon asymmetry of the universe.
\newblock {\em Phys. Lett. B} {620}, 17--26 (2005), arXiv:hep-ph/0505013.

\bibitem{Hambye2016}
T.~Hambye and D.~Teresi,
\newblock Higgs doublet decay as the origin of the baryon asymmetry,
\newblock {\em Phys. Rev. Lett.} {117}, 091801 (2016), arXiv:1606.00017 [hep-ph].

\bibitem{Hambye2017}
T.~Hambye and D.~Teresi,
\newblock Baryogenesis from $L$-violating Higgs-doublet decay in the density-matrix formalism,
\newblock {\em Phys. Rev. D} {96}, 015031 (2017), arXiv:1705.00016 [hep-ph].

\bibitem{barry2010}
J.~Barry and W.~Rodejohann.
\newblock Lepton mixing and neutrino mass generation from $s_4$ symmetry.
\newblock {\em Phys. Rev. D} { 81}, 093002 (2010).

\bibitem{Ma2001}
E.~Ma.
\newblock Naturally small seesaw neutrino mass with no new physics beyond the {TeV} scale.
\newblock {\em Phys. Rev. Lett.} {86}, 2502--2504 (2001), arXiv:hep-ph/0011121.

\bibitem{Sarma2020}
L.~Sarma, P.~Das and M.~K.~Das,
\newblock Scalar dark matter and leptogenesis in the minimal scotogenic model,
\newblock {\em Nucl. Phys. B} {963}, 115300 (2021), arXiv:2004.13762 [hep-ph].

\bibitem{Borah2019}
D.~Borah, P.~B. Dev, and A.~Kumar.
\newblock Tev scale leptogenesis, inflaton dark matter and neutrino mass in a scotogenic model.
\newblock {\em Phys. Rev. D} {99}, 055012 (2019), arXiv:1810.03645 [hep-ph].

\bibitem{Harrison:2002er}
P.~F.~Harrison, D.~H.~Perkins and W.~G.~Scott,
\newblock Tri-bimaximal mixing and the neutrino oscillation data,
\newblock {\em Phys. Lett. B} {530}, 167--173 (2002), arXiv:hep-ph/0202074.

\bibitem{An:2012eh}
Daya Bay collaboration (F.~P.~An {\it et al.}),
\newblock Observation of electron-antineutrino disappearance at Daya Bay,
\newblock {\em Phys. Rev. Lett.} {108}, 171803 (2012), arXiv:1203.1669 [hep-ex].

\bibitem{Ahn:2012nd}
RENO collaboration (J.~K.~Ahn {\it et al.}),
\newblock Observation of reactor electron antineutrinos disappearance in the RENO experiment,
\newblock {\em Phys. Rev. Lett.} {108}, 191802 (2012), arXiv:1204.0626 [hep-ex].

\bibitem{Abe:2012tg}
Double Chooz collaboration (Y.~Abe {\it et al.}),
\newblock Indication of reactor $\bar{\nu}_e$ disappearance in the Double Chooz experiment,
\newblock {\em Phys. Rev. Lett.} {108}, 131801 (2012), arXiv:1112.6353 [hep-ex].

\bibitem{Luhn:2013fh}
C.~Luhn,
\newblock Trimaximal TM1 neutrino mixing in $S_4$ with spontaneous CP violation,
\newblock {\em Nucl. Phys. B} {875}, 80--100 (2013), arXiv:1306.2358 [hep-ph].

\bibitem{Grimus:2013pg}
W.~Grimus,
\newblock Discrete symmetries, roots of unity, and lepton mixing,
\newblock {\em J. Phys. G} { 40}, 075008 (2013), arXiv:1301.0495 [hep-ph].

\bibitem{Rodejohann:2012kx}
W.~Rodejohann and H.~Zhang,
\newblock Simple two parameter description of lepton mixing,
\newblock {\em Phys. Rev. D} { 86}, 093008 (2012), arXiv:1207.1225 [hep-ph].

\bibitem{Ganguly:2023jml}
J.~Ganguly, J.~Gluza, B.~Karmakar and S.~Mahapatra,
\newblock Phenomenology of the flavor symmetric scoto-seesaw model with dark matter and TM1 mixing,
\newblock {\em Phys. Rev. D} {110}, no. 3, 035012 (2024), arXiv:2311.15997 [hep-ph].

\bibitem{lam2008}
C.~S. Lam,
\newblock The unique horizontal symmetry of leptons,
\newblock {\em Phys. Rev. D} {78}, 073015 (2008), arXiv:0809.1185 [hep-ph].

\bibitem{altarelli2010}
G.~Altarelli and F.~Feruglio,
\newblock Discrete flavor symmetries and models of neutrino mixing,
\newblock {\em Rev. Mod. Phys.} { 82}, 2701--2729 (2010), arXiv:1002.0211 [hep-ph].

\bibitem{antusch2004}
S.~Antusch and S.~F. King,
\newblock Neutrino mixing from the tri-bimaximal neutrino mixing ansatz,
\newblock {\em Phys. Lett. B} {597}, 199--204 (2004).

\bibitem{Weinberg1979}
S.~Weinberg,
\newblock Baryon- and lepton-nonconserving processes,
\newblock {\em Phys. Rev. Lett.} {43}, 1566--1570 (1979).

\bibitem{Ma1998}
E.~Ma,
\newblock Pathways to naturally small neutrino masses,
\newblock {\em Phys. Rev. Lett.} {81}, 1171--1174 (1998).

\bibitem{Gunion:2002zf}
J.~F. Gunion and H.~E. Haber,
\newblock The CP conserving two Higgs doublet model: The Approach to the decoupling limit,
\newblock {\em Phys. Rev. D} { 67}, 075019 (2003).

\bibitem{DavidsonLogan2009}
S.~M. Davidson and H.~E. Logan,
\newblock Dirac neutrinos from a second Higgs doublet,
\newblock {\em Phys. Rev. D} {80}, 095008 (2009), arXiv:0906.3335 [hep-ph].

\bibitem{DUNE:2020lwj}
DUNE collaboration,
\newblock Long-baseline neutrino oscillation physics potential of the DUNE experiment,
\newblock {\em Eur. Phys. J. C} {80}, 978 (2020), arXiv:2006.16043 [hep-ex].

\bibitem{AliAjmi:2024xus}
A.~Ajmi,
\newblock Status of the Hyper-Kamiokande Experiment,
\newblock {\em PoS} HQL2023, 098 (2024).

\bibitem{Esteban:2024eli}
I.~Esteban {\it et al.},
\newblock NuFit-6.0: updated global analysis of three-flavor neutrino oscillations,
\newblock {\em JHEP} {12}, 216 (2024), arXiv:2410.05380 [hep-ph].

\bibitem{Planck2018}
Planck collaboration (N.~Aghanim {\it et al.}),
\newblock Planck 2018 results. VI. Cosmological parameters,
\newblock {\em Astron.\ Astrophys.} {641}, A6 (2020), arXiv:\allowbreak{}1807.06209 \allowbreak{}[astro-ph.CO].

\bibitem{Thapa:2023fxu}
B.~Thapa {\it et al.},
\newblock A minimal inverse seesaw model with $S_4$ flavour symmetry,
\newblock {\em JHEP} { 11}, 154 (2023), arXiv:2305.09306 [hep-ph].

\bibitem{KamLAND-Zen2016}
KamLAND-Zen collaboration,
\newblock Search for majorana neutrinos near the inverted mass hierarchy region with KamLAND-Zen,
\newblock {\em Phys. Rev. Lett.} { 117}, 082503 (2016), arXiv:1605.02889 [hep-ex].

\bibitem{GERDA2020}
GERDA collaboration,
\newblock Improved limit on neutrinoless double-$\beta$ decay of $^{76}$Ge from GERDA phase II,
\newblock {\em Phys. Rev. Lett.} { 125}, 252502 (2020), arXiv:1803.11100 [nucl-ex].

\bibitem{EXO-2002019}
EXO-200 collaboration,
\newblock Search for neutrinoless double-$\beta$ decay with the complete EXO-200 dataset,
\newblock {\em Phys. Rev. Lett.} {123}, 161802 (2019), arXiv:1906.02723 [hep-ex].

\bibitem{MEG:2016}
MEG collaboration (A.~Baldini {\it et al.}),
\newblock Search for the lepton flavour violating decay $\mu^+ \to e^+ \gamma$ with the full dataset of the MEG experiment,
\newblock {\em Eur. Phys. J. C} {76}, no. 8, 434 (2016), arXiv:1605.05081 [hep-ex].

\bibitem{Guo:2017}
C.~Guo {\it et al.},
\newblock Hunting for Heavy Majorana Neutrinos with Lepton Number Violating Signatures at LHC,
\newblock {\em JHEP} {04}, 065 (2017), arXiv:1701.02463 [hep-ph].

\bibitem{Rubakov:1996}
V.~A. Rubakov and M.~E. Shaposhnikov,
\newblock Electroweak baryon number nonconservation in the early universe and in high-energy collisions,
\newblock {\em Usp. Fiz. Nauk} {166}, 493--537 (1996), arXiv:hep-ph/9603208.

\bibitem{Davidson:2002}
S.~Davidson and A.~Ibarra,
\newblock A Lower bound on the right-handed neutrino mass from leptogenesis,
\newblock {\em Phys. Lett. B} { 535}, 25--32 (2002), arXiv:hep-ph/0202239.

\bibitem{Buchmuller:2002}
W.~Buchmuller, P.~Di~Bari and M.~Plumacher,
\newblock Cosmic microwave background, matter-antimatter asymmetry and neutrino masses,
\newblock {\em Nucl. Phys. B} { 643}, 367--390 (2002), arXiv:hep-ph/0205349.

\bibitem{Antusch:2011nz}
S.~Antusch {\it et al.},
\newblock Leptogenesis in the Two Right-Handed Neutrino Model Revisited,
\newblock {\em Phys. Rev. D} { 86}, 023516 (2012), arXiv:1107.6002 [hep-ph].

\bibitem{Hugle:2018qbw}
T.~Hugle, M.~Platscher and K.~Schmitz,
\newblock Low-Scale Leptogenesis in the Scotogenic Neutrino Mass Model,
\newblock {\em Phys. Rev. D} { 98}, no. 2, 023020 (2018), arXiv:1804.09660 [hep-ph].

\bibitem{2016}
Planck collaboration (P.A.R.~Ade {\it et al.}),
\newblock Planck 2015 results: XIII. Cosmological parameters,
\newblock {\em Astron.\ Astrophys.} {594}, A13 (2016), arXiv:\allowbreak{}1502.01589 \allowbreak{}[astro-ph.CO].


\bibitem{Sarma:2022qka}
L.~Sarma, P.~K. Paul and M.~K. Das,
\newblock Connecting dark matter, baryogenesis and neutrinoless double beta decay in a $A_4 \otimes Z_8$-based $\nu$2HDM,
\newblock {\em Int. J. Mod. Phys. A} { 37}, no. 27, 2250157 (2022),  	arXiv:2208.14764 [hep-ph].


\bibitem{Zhao:2011}
Z.-H. Zhao,
\newblock {Realizing tri-bimaximal mixing in minimal seesaw model with $S_4$ family symmetry},
\newblock {\em Phys. Lett. B} { 701} (2011) 609--613.

\bibitem{collaboration2021legend1000}
LEGEND Collaboration, N.~Abgrall, I.~Abt, et~al.
\newblock {LEGEND-1000 Preconceptual Design Report}.
\newblock 2021.
\newblock arXiv:2107.11462 [physics.ins-det].

\bibitem{Adhikari_2021}
G.~Adhikari, S.~Al Kharusi, E.~Angelico, et~al.
\newblock {nEXO: neutrinoless double beta decay search beyond $10^{28}$ year half-life sensitivity}.
\newblock \emph{Journal of Physics G: Nuclear and Particle Physics}, 49(1):015104, 2021.
\newblock ISSN 1361-6471.
\newblock DOI: \href{https://doi.org/10.1088/1361-6471/ac3631}{10.1088/1361-6471/ac3631}.

\end{thebibliography}






\end{document}